\documentclass[rmp,aps,preprint,nofootinbib,groupedaddress,draft]{revtex4-1}

\usepackage{amsfonts,amsmath,amssymb}
\usepackage{mathrsfs}%
\DeclareSymbolFontAlphabet{\mathcop}{rsfs}
\usepackage[mathscr]{eucal}
\usepackage{color}
\usepackage{graphicx}

\graphicspath{{imma/}}

\newcommand{\be}{\begin{equation}}
\newcommand{\ee}{\end{equation}}

\newcommand{\stre}[2]{\genfrac{[}{]}{0pt}{0}{#1}{#2}}%
\newcommand{\tstre}[2]{\genfrac{[}{]}{0pt}{1}{#1}{#2}}%
\DeclareMathOperator{\rank}{rank}
\usepackage{lineno,hyperref}
\modulolinenumbers[5]

\begin{document}


\title{Higher-order theories of gravity: diagnosis, extraction and reformulation via non-metric extra
degrees of freedom}

\author{Alessio Belenchia}
\email[E-mail address: ]{alessio.belenchia@oeaw.ac.at}

\affiliation{Institute for Quantum Optics and Quantum Information (IQOQI), Boltzmanngasse 3 1090 Vienna, Austria.}

\author{Marco Letizia}
\email[E-mail address: ]{mletizia@sissa.it}

\author{Stefano Liberati}
\email[E-mail address: ]{liberati@sissa.it}

\affiliation{SISSA --- International School for Advanced Studies, Via Bonomea 265, 34136 Trieste, Italy.}
\affiliation{INFN Sez. di Trieste, via Valerio 2, 34127, Trieste, Italy.}

\author{Eolo Di Casola}
\email[E-mail address: ]{dicasola@sissa.it}

\affiliation{SISSA --- International School for Advanced Studies, Via Bonomea 265, 34136 Trieste, Italy}

\begin{abstract}
Modifications of Einstein's theory of gravitation have been extensively considered in the past years, in connection to both cosmology and quantum gravity. Higher-curvature and higher-derivative gravity theories constitute the main examples of such modifications. These theories exhibit, in general, more degrees of freedom than those found in standard General Relativity; counting, identifying, and retrieving the description/representation of such dynamical variables is currently an open problem, and a decidedly nontrivial one. In this work we review, via both formal arguments and custom-made examples, the most relevant methods to unveil the gravitational degrees of freedom of a given model, discussing the merits, subtleties and pitfalls of the various approaches.
\end{abstract}





\maketitle

\tableofcontents

\clearpage

\section{Introduction \label{intro}}

\subsection{General relativity and beyond: a bird's eye view}

The theory of General Relativity (GR) is now exactly one century old, yet it seems to enjoy a never-ending spring. Its connection with observations and experiments is probably stronger than ever~\cite{Abbott:2016blz,Abbott:2016nmj,Aasi:2013wya}, and new theoretical branches are blossoming at regular pace.

Still, the innermost nature of gravity remains as yet an unsolved riddle. Our understanding of large-scale structures leads to puzzling conclusions~\cite{Bertone:2004pz,Olive:2003iq,Lukovic:2014vma}, the very early phases of the Cosmos demand a clearer picture~\cite{Weinberg:2005vy,Vilenkin:1987kf}, and the marriage between the micro-world and the macroscopic picture is ``perfectly \emph{un}happy''~\cite{DeWitt:2007mi,Oriti:2009zz}. Even in the narrower context of classical (i.e., non-quantum) gravity, we ought to admit that the picture is still somehow blurred~\cite{Mathur:2009hf,Giddings:1995gd,Misner:1965zz,Godel:1949ga}. All in all, as Newton puts it, we are stuck on a shore, ``whilst the great ocean of truth lay all undiscovered before [us]''~\cite[Chap.~27]{brewsternewton}.

To face this challenge, the theoretical landscape has immensely  expanded, and many competing models have surfaced, ranging from the phenomenologically viable to the decidedly speculative. Stripped to the bone, the common goal remains the same: identifying the best possible description for gravitational phenomena, by framing the most ``correct'' representation for the degrees of freedom (d.o.f.'s), and their dynamics, while remaining faithful to the observational and experimental contraints. The way this goal is achieved, however, greatly varies within the broad spectrum of what are now known as the \emph{generalised/alternative/extended theories of gravity} (ETG's).

In most of the cases, the ETG's are formulated in such a way that a specific geometric interpretation is associated to (at least some of) the elements playing a role in the gravitational action and/or field equations, mirroring the widely accepted chronogeometric meaning of the tensor $g_{\mu\nu}$ in Einstein's model~\cite{will-rel,misner-relgen,Wald:1984rg}. The geometric structures may or may not be dynamical themselves, even though they typically experience a form of evolution. What instead is usually left fixed once and for all, are the signature and topology of the spacetime itself, but of course exceptions have been conceived~\cite{Odintsov:1994ge}.

Apart from these fundamental common grounds, the intricate jungle of ETG's offers any sort of variation on the given theme~\cite{Clifton:2011jh}. There are metric and non-metric theories, together with metric-affine, affine, and purely affine proposals; gravitational degrees of freedom of any nature (scalar, vectorial, tensorial, spinorial, etc.) are juxtaposed to the usual graviton, and field equations of arbitrarily high order can be easily conceived --- even full non-local models are currently under scrutiny~\cite{ArkaniHamed:2002fu}. Violations of the various equivalence principles are allowed, both in the gravitational and matter sector; and candidate theories in higher and lower spacetime dimensions have been advanced~\cite{Padmanabhan:2010zzb}, motivated by AdS/CFT correspondence~\cite{Klebanov:2000me} and other daring conjectures.

The catalogue is vast and variously interwoven, and its review immediately generates a few key questions, one of which is the focus of the following pages.

\subsection{ETG's in a nutshell, and ``the problem''}

In a very large sample of the set of ETG's, a typical element is the presence of gravitational d.o.f.'s encoded in objects other than the mere metric field $g_{\mu\nu}$. The specific choice of the added building blocks, and of their couplings with the metric, is what determines the subsequent structure of the theory. We can thus invent scalar-tensor theories, vector-tensor theories, scalar-vector-tensor, or multi-scalar-tensor theories of different flavours (see e.g. Refs~\cite{Sotiriou:2006hs,Moffat:2005si}).

Another road taken to introduce new variables is that of enlarging the geometric structure available on the bare, underlying manifold. At the very least, one can pick an affine structure which is independent from the metric one; the connection coefficients thus become new dynamical variables (as done in the Palatini method of variation, and in all the subset of affine, purely affine, and metric-affine theories, see e.g. Refs~\cite{Hehl:1994ue,Vitagliano:2010sr,Sotiriou:2006qn}). A similar standpoint requires to consider the role of torsion~\cite{Szczyrba:1985pd,Arcos:2005ec} and/or non-metricity~\cite{Vitagliano:2013rna}, up to the farthest consequences (e.g. Weintzb{\"o}ck teleparallelism~\cite{Aldrovandi:2013wha,Obukhov:2002tm}).

Finally, and this is the main topic of the present work, a class of ETG's can be built by focussing exclusively on the dynamics of the metric, but allowing for a more complex form of the gravitational action. The resulting theories exhibit, as we shall see in detail below, a larger number of metric gravitational d.o.f.'s.

This last, special sub-class of ETG's can be referred to as \emph{higher-curvature theories} (from the form of the associated actions, see Sect~\ref{history}), or \emph{higher-derivative theories} (from the structure of the resulting field equations). While the two expressions widely overlap and are often used interchangeably, there is indeed a crucial difference: a higher-curvature ETG can nonetheless result in differential field equations of order two as in GR --- this occurs for the whole class of Lanczos--Lovelock models~\cite{Padmanabhan:2010zzb,Padmanabhan:2013xyr} --- whereas a higher-derivative action will always give higher-order field equations. At any rate, when this difference is of little importance, the umbrella-term \emph{higher-order theories} can be used, to encompass at once all the possible cases.

Whichever way one looks at the landscape of ETG's, it ought to be remembered that, in the end, we aim for the correct number of gravitational d.o.f.'s, and their actual dynamics; the rest is a matter of representations and mathematical rearrangements of variables into suitably interpreted geometric quantities. Such a conclusion naturally leads to a relevant problem.

Suppose to have two ETG's emerging from different subsets of the catalogue, and treating gravitational phenomena in very different ways. If the number of gravitational d.o.f.'s turns out to be the same, and also their dynamics qualifies as the same, then it may be fair to say that we are not looking at two different theories anymore: rather, we are just dealing with two facets of the very \emph{same} description of reality. Once the ``dictionary'' bridging the two models becomes available, there is no need anymore for two paradigms: we are merely considering two separate \emph{representations} of the same physical system, both carrying essentially the same semantic content.

Extracting and comparing the actual dynamics of all the ETG's --- especially those for which the ``true'' variables not manifest --- is then a crucial step towards unveiling the hidden network of relations (and redundancies) within the catalogue. Also, it can be relevant for assessing the level of actual originality in a newly-proposed model, or its potential to explain the puzzling observational features of our Universe. Finally, this might even be considered a --- humble, preliminary, tentative --- first step towards a ``meta-theory'' of gravitation~\cite{Sotiriou:2007zu}.  

Yet, achieving this goal is a highly non-trivial task. Many techniques are available, which try to pin down the d.o.f.'s, each one presenting its own pros and cons, with a plethora of pitfalls and subtleties that make it hard to build a universal protocol. Even worse, some methods are heavily background-dependent, or they result in macroscopic modifications of the actions and field equations, to the point that the comparison between paradigms can become almost meaningless.

Still, given the great relevance of the extraction of the dynamical content of any ETG, we deem it useful --- or rather, necessary --- to critically review the available protocols, and highlight what might be the best way to deal with this crucial issue in gravitational theory.

\subsection{This article at a glance}

The internal organisation of this work goes as follows. In the next Section~\ref{prelim} we shall properly frame the context of our discussion, and sum up some results about the gravitational d.o.f.'s and their various representations. There will be room to highlight the crucial role of the \emph{boundary terms}, and explore the opportunities of a simple diagnostic tool based on surface counter-terms.

Section~\ref{propagators} hosts a detailed analysis of the \emph{linearisation procedure}, and of its main advantages and dangers as per the extraction of number and nature of the gravitational d.o.f.'s. We shall dive into the momentum representation and its subtleties, and focus on the notion of a \emph{propagator} in ETG's --- as long as such concept makes sense --- and on its application to higher-order theories.

Section~\ref{Auxiliary fields} will shift the attention onto the alternative route of the use of \emph{auxiliary fields}, outlining its pros and cons, and discussing the limits of this admittedly powerful technique. Some aptly crafted cases will prove how delicate and intricate is the choice of a suitable set of alternative variables, and how the latter affects the dynamical structure.

Section~\ref{linearization} is devoted to an extraction method for the d.o.f.'s which has the appearance of a ``master key'', and works effectively wherever the \emph{expansion around a maximally symmetric spacetime} is an allowed procedure. The protocol is indeed powerful, but it has its well-hidden pitfalls, which we shall discuss with the aid of a few custom-made examples highlighting the most subtle aspects.

Finally, Section~\ref{hamiltonian} is dedicated to the top-level diagnostic tool for the gravitational dof's, i.e. the \emph{Hamiltonian formalism}. We shall review the extension of the method to higher-order ETG's, address its grey areas and potential risks, and devote a few paragraphs to miscellaneous remarks on more exotic, yet noteworthy pathways (peaking with the concept of Einstenian ``strength'' of a system of field equations).

The last paragraphs (Sect.~\ref{conclusions}) will be devoted to the main conclusions and to some speculations about possible future developments.

\subsubsection{Relevant notations and conventions \label{notat}}

The literature on the topic of ETG's is vast and diverse, and its contributions come from many communities, each one sporting a set of its own standards. In this work, we have tried to harmonise notations as much as possible.

For basic conventions, we mostly follow the authoritative Refs.~\cite{misner-relgen,Wald:1984rg}. Spacetime is assumed to be a $4$-dimensional manifold $\mathcal{M}$, equipped with a pseudo-Riemannian metric $g_{\mu\nu}$ with Lorentzian signature. In an instantaneous free-fall reference frame it is $g_{\mu\nu} = \mathrm{diag} (-1,1,1,1)$. The connection coefficients are $\Gamma^\lambda_{\mu\nu}$, and the Riemann curvature tensor is defined as ${R^{\alpha}}_{\beta\gamma\delta} := \partial_\gamma \Gamma^\alpha_{\beta\delta} - \partial_\delta \Gamma^\alpha_{\beta\gamma} + \dots$. The Ricci tensor is the contraction $R_{\mu\nu} := {R^{\lambda}}_{\mu\lambda\nu}$, and the scalar curvature is the trace $R := g^{\mu\nu} R_{\mu\nu}$. The Einstein tensor reads $G_{\mu\nu} := R_{\mu\nu} - g_{\mu\nu} R /2$. The d'Alembertian derivative operator is denoted as $\square := \nabla_{\! \alpha} \nabla^\alpha$.

As for other specific notations: by $S_{\mu\nu}$ and $C_{\alpha\beta\gamma\delta}$ we denote the trace-free part of the Ricci tensor, and the fully traceless Weyl conformal tensor, respectively. The script letter $\mathcal{G}$ is used for the Gauss--Bonnet quadratic combination, $\mathcal{G} := R^2 - 4 R^{\mu\nu} R_{\mu\nu} + R^{\alpha\beta\gamma\delta} R_{\alpha\beta\gamma\delta}$.

Finally, a note on the terminology used about ETG's: as emphasized, we shall refer to a \emph{higher-derivative theory} when the resulting field equations turn out to exhibit differential order higher than two, whereas we shall refer to a \emph{higher-curvature theory} when the action is expressed in terms of combinations of the curvature tensor other than the standard Einstein-Hilbert term. The umbrella-term \emph{higher-order theory} will be found in reference to both types, indistinguishably.
%
\section{State-of-the-art and some relevant technicalities \label{prelim}}

\subsection{Emerging patterns in higher-order ETG's \label{history}}

Einstein's GR is known to be a \emph{purely metric} theory of gravity~\cite{will-rel}, which stems from the variation of Hilbert's action~\cite{misner-relgen,Wald:1984rg}. This means that the common origin of all gravitational phenomena is attributed exclusively to the dynamics of the metric tensor $g_{\mu\nu}$, hosted on a 4-dimensional manifold $\mathcal{M}$. The starting point is the action
\begin{equation}
\mathcal{S}_{\text{E.H.}} = \frac{1}{2 \kappa} \int_{\mathcal{M}} R \, \sqrt{-g} \, \mathrm{d}^4 x \;, \label{eq.ehaction}
\end{equation}
with $\kappa$ a coupling constant, and $R$ the scalar curvature. Once the above action is varied with respect to $g_{\mu\nu}$ (or, for computational convenience, to $g^{\mu\nu}$), the field equations emerge
\begin{equation}
G_{\mu\nu} + \Lambda g_{\mu\nu} = 0 \;, \label{eq.grfieldeqs}
\end{equation}
with $\Lambda$ a fundamental constant, and $G_{\mu\nu}$ the symmetric, divergence-free Einstein tensor. The system of second-order, quasi-linear PDE's in Eq.~\eqref{eq.grfieldeqs} above allows one to determine the gravitational configurations of spacetime.

Two brief remarks: the above formulation of GR does not pin down the \emph{actual} number of gravitational d.o.f.'s --- which is way smaller than the ten free components of the symmetric tensor $g_{\mu\nu}$, see Sect.~\ref{gravdofs} ---, nor it justifies completely the emergence of the field equations from action~\eqref{eq.ehaction}, the problem being the lack of the correct boundary terms (see Sect.~\ref{boundaryterms}). After a quick fixing, however, the formulation becomes robust, and GR works (almost) flawlessly.

The class of higher-order ETG's builds upon the same premise behind Einstein's model, i.e. it encompasses a wide range of (supposedly) purely metric theories of gravity. The main difference lies in the shape of the starting action, which admits arbitrarily complex contributions from the curvature tensor. In all generality, Hilbert's action is then traded for something like --- see~\cite{Brown:1995su}
\begin{equation}\label{genaction}
\mathcal{S} = \frac{1}{2 \kappa}\int_\mathcal{M} \sqrt{-g}\,\mathrm{d}^4 x \, f\left(g_{\mu\nu},R_{\mu\nu\rho\sigma},\nabla_{\alpha_1} R_{\mu\nu\rho\sigma},...,\nabla_{(\alpha_1}...\nabla_{\alpha_m)}R_{\mu\nu\rho\sigma}\right).
\end{equation} 
with $f$ an aptly defined scalar function of the Riemann tensor, its covariant derivatives, and the metric tensor,\footnote{Typically, the function $f$ is chosen such as to be analytic in its arguments, so that it admits a Taylor-series expansion in the fileds $g_{\mu\nu}$ or $R_{\alpha\beta\gamma\delta}$ allowing one to estimate some physical effects at different orders. Quadratic corrections are quite common in the literature, whereas anything beyond order-$2$ is often studied in connection with simpler actions, as in $R^n$-theories, to keep the calculations manageable.} and $\kappa'$ the accordingly-redefined coupling constant. The dynamical variables are still encoded in $g_{\mu\nu}$, but the change in the action functional affects the shape of the field equations. 

We can thus expect that higher-curvature contributions will result in higher-derivative operators acting on $g_{\mu\nu}$, with the onset of field equations of order higher than two, e.g., an action constructed with $f \left( g_{\mu\nu} , {R^\alpha}_{\beta\gamma\delta} \right)$ generates fourth-order field equations. This rise in the differential character of the equations of motion is a fairly general result, even though notable exceptions exist: in 4 spacetime dimensions, the quadratic Gauss--Bonnet theory actually gives back second-order only field equations, despite being a higher-curvature ETG~\cite{Padmanabhan:2010zzb}. Also, as soon as one allows for the introduction of higher differential operators in the action, such as in the case of $f \left( g_{\mu\nu} , \square {R^\alpha}_{\beta\gamma\delta} \right)$-theories and models alike, the field equations will ramp up accordingly.

That gravity may be described by incorporating higher contributions from the curvature is an idea one can accept based on various grounds, from observational needs to purely theoretical motivations. For instance, in semiclassical treatments of gravity most of the tentative quantisation processes require the introduction of higher-order corrections to deliver a renormalizable theory~\cite{Capozziello:2010zz}. At the bare minimum, a field expansion demands models with quadratic contributions from the curvature. Also, one might adopt some sort of ``effective'' standpoint by incorporating our current lack of understanding of the quintessential dynamics of gravity into non-local interactions --- and hence, into a ladder of truncated higher-order expansions~\cite{Weinberg:1980gg}. Finally, there are purely formal reasons to go beyond Hilbert's action, as the ``simple'' Eq.~\eqref{eq.ehaction} might be just the lowest step in a ladder of increasing complexity, and this mathematical structures alone might be worthy some investigation. At any rate, higher-order ETG's are now a familiar character in the landscape of gravitation models, and as such they deserve to be properly understood, also in their taxonomic relation to the other branches of the ``family tree'' of competing models sketched above.

Basic examples of such theories are the $f (g_{\mu\nu} , R)$ theories --- more commonly known as $f (R)$-theories~\cite{DeFelice:2010aj,Sotiriou:2008rp,Amendola:2006we}. There, the instances of the curvature appear only via the fully traced scalar $R$. Fairly ubiquitous in cosmological contexts, these ETG's remain among the most popular to provide alternative explanations to large-scale phenomena.

Then, we have the the $f (g_{\mu\nu} , R_{\mu\nu})$, where the Ricci tensor is allowed as well to enter the game, often to comply with the introduction of semiclassical corrections of quantum nature~\cite{Allemandi:2004wn,Soussa:2003re}.

Finally, one can conceive the completely general $f (g_{\mu\nu} , {R^\alpha}_{\beta\gamma\delta})$, where the whole Riemann tensor is allowed to contribute to the action~\cite{Deruelle:2009zk}. Within this sub-class, a well-known specimen is Weyl's conformal gravity, where only the traceless part of the curvature enters the Lagrangian density.

The group of higher-order ETG's can then be further expanded by accepting actions where not only one has non-trivial combinations of the curvature tensor, but also covariant derivatives of the Riemann tensor, such as in the class of $f (g_{\mu\nu} , {R^\alpha}_{\beta\gamma\delta} , \nabla_\sigma {R^\alpha}_{\beta\gamma\delta})$-theories and so forth.

The rise of higher-order ETG's naturally suggests the question: what could be a way to interpret the convoluted dynamics of such models? Can we devise a protocol to transform higher-curvature actions (and, hence, higher-derivative field equations) into dynamically equivalent models with a more familiar, second-order evolution? The answer to this question is positive, at least for a large number of higher-order ETG's, which can be remapped into second-order theories for the metric (now sporting fewer actual d.o.f.'s), plus other dynamical objects --- scalars, vectors, tensors, spinors, as we shall see in the following.

Such possibility, however, posits another important question: are the higher-order ETG's truly ``purely metric'' as initially stated? Or are they just other type of ETG's in disguise, where the non-metric variables are forcibly incorporated into $g_{\mu\nu}$? And what is the true meaning of the expression ``purely metric'', if any?

To date, a \emph{general} answer to all such questions is not known, nor is a (manageable) standard to perform such transformations in any possible case. What we have is a crowded toolbox of scattered and often \emph{ad hoc} recipes dictating how to count and frame the actual gravitational d.o.f.'s of a given model, but a careful inspection of these protocols shows pitfalls and subtleties behind any corner.

A more honest (re)starting point, then, might be first and foremost a detailed review of the available techniques, leaving no grey areas behind, and providing a reliable picture of the available results from the very ground up.

\subsection{Gravitational degrees of freedom \label{gravdofs}}

As a fair beginning for our discussion, we deem it necessary to give here a brief introduction to the matter of gravitational d.o.f.'s, starting from the very terminology.

By ``degree of freedom'' we mean here any dynamical variable involved in the actual unfolding of a physical system. In a Lagrangian or Hamiltonian description of a physical system, the degrees of freedom are the dynamical entities which fully describe the evolution of the system once all the constraints have been considered.

An important remark is needed at this stage: while the mentioned degrees of freedom are the sole elements encoding the actual physical behaviour of a system, and the ones sought-after to produce a model or a theory for some aspects of Nature, it is often useful to classify them according to their properties under certain symmetry transformations. Since all Lorentzian manifolds are locally described by Minkowski spacetime, it is natural to classify the degrees of freedom according to the symmetries of flat spacetime i.e., according to the irreducible representations of the Poincar\'{e} group. This can be done either expanding the theory around flat spacetime, or \textit{covariantizing} a certain set of conditions that are germane to fields of spin $s$ in flat spacetime (e.g., the covariant Fierz--Pauli condition for spin-$2$ fields \cite{Hindawi:1995an}). However, if one is interested in specific non-flat backgrounds, the classification of the degrees of freedom can be associated to different symmetry groups. In particular in cosmology, the d.o.f.'s are usually classified according to their helicity under spatial rotations.

The issue with the degrees of freedom in a field theory is then twofold: on one hand, we are asked to find \emph{how many} dynamical variables are there (together with, of course, their specific dynamical character); on the other hand, we have to establish their \emph{formal character}, i.e. a given representation of the theory, in terms of a specific sets of fields and/or geometric quantities.

When it comes to gravity, the case with the degrees of freedom gets slightly tricky, and it is crucial to correctly understand the logic. Consider, to begin with, the field equations in Eq.~\eqref{eq.grfieldeqs} with $\Lambda = 0$. Taken as is, the expression $G_{\mu\nu} = 0$ stands for a system of second-order, quasilinear, hyperbolic partial differential equations governing the dynamics of the symmetric, rank-$2$ tensor, $g_{\mu\nu}$. The gravitational degrees of freedom, whatever they might be, must then be encoded in $g_{\mu\nu}$.

If we represent the metric tensor by a $4 \times 4$ matrix in an arbitrary coordinate system, its fundamental symmetry property $g_{\mu\nu} = g_{\nu\mu}$ implies that we are actually working with \emph{ten} independent components, instead of the full sixteen available in a generic rank-$2$, 4-D tensor. We then have to take into account the background independence of the field equations and of the underlying Lagrangian, i.e. the fact that the points on the manifold $\mathcal{M}$ can be arbitrarily relabelled without affecting the dynamics. This implies that, of the ten gravitational variables, \emph{four} are unphysical (expression of the freedom in redefining the coordinate system), and the actual dynamics is encoded only in the remaining \emph{six} free parameters.

On top of that, we have also to consider the role of the field equations. The easiest way to see what happens is to pick an apt foliation of the spacetime manifold into spatial leaves evolving along the streamlines of an affine parameter (a ``time'' variable), and project the field equations on this stack of $3$-spaces dynamically evolving --- this is tantamount to selecting a special set of coordinates (ADM formalism, see Sect.~\ref{hamiltonian}). It can be shown~\cite{Choquet-Bruhat:2009xil} that \emph{four} out of the ten field equations are just constraint equations, providing no contribution to the actual dynamics of gravity; this affects the number counting of the degrees of freedom, which gets ultimately reduced to the final figure of \emph{two}.\footnote{For a decidedly different technique leading to the same result, see Sect.~\ref{forza}.}

Suppose now to introduce a weak-field approximation, so that the metric $g_{\mu\nu}$ can be decomposed into the sum of a background, flat part $\eta_{\mu\nu}$, and a small perturbation $\varepsilon \, h_{\mu\nu}$ (with $\varepsilon$ a bookkeeping parameter accounting for the order in a Taylor-series expansion of the field). The presence of the Minkowskian background allows for the introduction of a further, global symmetry, the Poincar{\'e} invariance lying beneath the laws of Special Relativity.

The gravitational field embodied by $\varepsilon \, h_{\mu\nu}$ sports two degrees of freedom, as we know from above; at the same rate, the perturbation tensor $\varepsilon \, h_{\mu\nu}$ can be classified according to the irreducible spinor representations of the Poincar{\'e} group acting on the flat background. It turns out that the two gravitational degrees of freedom can be rearranged into a spin-$2$ object living on Minkowski spacetime.

We thus conclude that, according to GR, the gravitational phenomena can be mediated by a spin-$2$ boson which, in view of the long-range character of gravitational interactions, must have vanishing mass: such mediator is commonly known as the \emph{graviton}.

The bottom line is thus that GR has only two gravitational d.o.f.'s, which can be characterised as the components of a massless, spin-$2$ graviton living on Minkowski --- or (anti-) de Sitter spacetime.\footnote{This last conclusion holds in view of the existence, on (anti-) de Sitter universes, of global symmetry groups analogue to the Poincar{\'e} group acting on Minkowski spacetime --- the so-called de Sitter--Fantappi{\'e}--Arcidiacono groups~\cite{Aldrovandi:2006vr,Benedetto:2009zza}. The irreducible representations of such groups allow to define the equivalent of integer and half-integer spinors (bosons and fermions), whence a resulting classification for ``particles''.}

What might be the analogous of such a conclusion for the vast class of higher-order ETG's is the main topic of the following pages.

\subsection{Boundary terms as diagnostic tools for the d.o.f.'s \label{boundaryterms}}

The variation of the Hilbert Lagrangian~\eqref{eq.ehaction} actually leads to Einstein's field equations only if one adds at least one of the two following, crucial assumptions: i) the manifold $\mathcal{M}$ over which the integrations in Eq.~\eqref{eq.ehaction} are performed is assumed to have a compact topology; ii) the variations $\delta \Gamma^\rho_{\sigma\tau}$ of the connection coefficients are assumed to vanish on the boundary $\partial \mathcal{M}$, together with the variations $\delta g_{\mu\nu}$ of the metric field.

As soon as these assumptions are relaxed --- as it often happens in a more general treatment of gravitational phenomena --- the mere variation of Hilbert's Lagrangian does not work anymore, and must be supplemented by additional terms if one wants to recover the field equations~\eqref{eq.grfieldeqs}.

To see this, let us go back to the variation of gravitational action~\eqref{eq.ehaction}, and perform it once again without any preemptive assumption on what occurs on the boundary $\partial \mathcal{M}$ (boundary which we take to be a generically non-compact structure). The outcome is, after some manipulations and integrations by part,
\begin{equation}
\begin{split}
\delta \mathcal{S} [g_{\mu\nu}] = &\int_{\mathcal{M}} G_{\mu\nu} \delta g^{\mu\nu} \, \sqrt{-g} \, \mathrm{d}^4 x  \\
&+ \int_{\mathcal{M}} \left( \delta^\alpha_\gamma \nabla_{\! \delta} g^{\delta\beta} - \nabla_{\! \gamma} g^{\alpha\beta} \right) \delta \Gamma^\gamma_{\alpha\beta} + \nabla_{\! \gamma} \left( g^{\alpha\beta} \delta \Gamma^\gamma_{\alpha\beta} - g^{\gamma\beta} \delta \Gamma^\alpha_{\alpha\beta} \right) \, \sqrt{-g} \, \mathrm{d}^4 x\;.
\end{split} \label{eq.ehvariation}
\end{equation}
The pieces on the second line need now be erased in some way, if one wants to recover Einstein's equations. The first term drops out in view of the metric compatibility condition $\nabla_{\! \alpha} g_{\mu\nu} = 0$, and so we are left with the last one, which vanishes if and only if we also assume $\delta \Gamma^\rho_{\sigma\tau} = 0$ on the boundary (or if we collapse the topology of $\mathcal{M}$ on a compact model).

A closer look at such two hypotheses shows, however, that they are indeed too restrictive. A compact topology is quite a peculiar configuration, and there is no reason to prefer it a priori over any other possible arrangement. In the same fashion, requiring the vanishing not only of the $\delta g_{\mu\nu}$'s on the boundary of $\mathcal{M}$, but also of their first derivatives there (it is $\delta \Gamma^\rho_{\sigma\tau} \sim \delta \partial_\rho g_{\sigma\tau}$), restricts too much the allowed set of field configurations, and ought to be avoided. 

This last issue becomes particularly annoying when one moves from GR to any higher-order ETG. Let us consider a scheme for which the field equations are of order $r$ in the derivatives of the metric, with $r > 2$. If we want to remove the additional derivative pieces in the action, the procedure sketched above demands the vanishing on $\partial \mathcal{M}$ of the variations of all the derivatives $\delta \partial^{(k)}_\lambda g_{\mu\nu}$ with $k = \left\{1, 2, \dots, r-1\right\}$. This requirement, however, affects the solutions of the field equations, as the $g_{\mu\nu}$'s have to comply with an additional set of derivative constraints, introduced just to render the variational problem well-posed, but without any link to the actual dynamics of gravity (the constraints are set long before the field equations are retrieved, let alone solved). As a result, the space of possible solutions gets reduced significantly, yet without any intervention of the field equations. Admittedly, this is too restrictive a condition, and should be traded for the single requirement of $\delta g_{\mu\nu} = 0$ on the boundary, without any further constraint.\footnote{Note the subtlety: when looking for the actual different solutions of the field equations, the set of initial data specified e.g. on a Cauchy surface must fix the values of the field and its derivatives up to order $(r-1)$ for the initial-value problem to be meaningful. At this stage, however, we are not dealing with \emph{single solutions} of the field equations, but rather with \emph{the space of all possible solutions}, as a whole. While the single-field configurations for a specified matter-energy distribution had rather be pinned down by the initial data, the space of admissible configurations emerging from the action is instead expected to be as large as possible, not to rule out any legitimate candidate.}

At least in the case of GR, the remaining terms in Eq.~\eqref{eq.ehvariation} can be reabsorbed successfully, as they turn out to be the variation themselves of twice the trace of the extrinsic curvature $K$ of the sub-manifold $\partial \mathcal{M}$.\footnote{Let the hypersurface $\partial \mathcal{M}$ be everywhere identified by the direction of its unit normal vector $n^\alpha$ (typically, one picks a spacelike boundary, and hence a timelike vector $n^\alpha$, but the results carry over to null boundaries as well). Then, the metric tensor $g_{\mu\nu}$ can be decomposed as $g_{\mu\nu} = \gamma_{\mu\nu} - n_\mu n_\nu$, with $\gamma_{\mu\nu}$ the metric tensor on $\partial \mathcal{M}$. Then, the extrinsic curvature is the tensor $K_{\mu\nu} := \nabla_{\! (\mu} n_{\nu)}$, and its trace is $K = g^{\mu\nu} K_{\mu\nu}$.} The Einstein--Hilbert action can thus be complemented by the additional surface integral
\begin{equation}
\mathcal{S}_{\text{G.H.Y.}} = 2 \oint_{\partial \mathcal{M}} K \, \sqrt{\gamma} \, \mathrm{d}^3 x \;, \label{eq.ghycountterm}
\end{equation}
known as the Gibbons--Hawking--York counter-term~\cite{Gibbons:1976ue,York:1972sj}, with $\gamma$ the determinant of the induced three-metric $\gamma_{ab}$. It is then only the full action given by
\begin{equation}
\mathcal{S}_{\text{grav}} = \frac{1}{2\kappa} \int_{\mathcal{M}} R \, \sqrt{-g} \, \mathrm{d}^4 x - \frac{1}{\kappa} \oint_{\partial \mathcal{M}} K \, \sqrt{\gamma} \, \mathrm{d}^3 x \;, \label{eq.gractionfull}
\end{equation}
which correctly delivers the set of Einstein's equations and nothing else, in any possible topological arrangement for the manifold $\mathcal{M}$, and with the sole requirement of the vanishing of the $\delta g_{\mu\nu}$'s on $\partial \mathcal{M}$.

\subsubsection{Boundary terms in ETG's}

The problem with the boundary terms just highlighted resurfaces whenever one considers a higher-order ETG. Three main aspects need be taken care of: i) the existence of the boundary terms; ii) their ability to erase all the uncompensated variations in the higher derivatives of the metric, and; iii) their use as a diagnostic tool suggesting the nature and number of the actual d.o.f.'s for the given model.

As for the first two (largely interwoven) issues, no general result or theorem is available to our knowledge, neither in a positive form, nor in that of a no-go statement. A wide range of partial results can be found in the literature, describing specific fixings of given actions, yet a full proposition is out of sight, perhaps because the number of possible variations on the GR theme is too wide. And even when boundary terms become available, they cannot account for all the uncompensated variations in the action: other bits must be turned off, or added in, entirely by hand.\footnote{A few ``lucky'' cases exist, though. For instance, Lanczos--Lovelock gravity (the class of $n$-dimensional generalisations of Gauss--Bonnet theory) is such that all the uncompensated terms can be accounted for by variations of surface terms generalising the Gibbons--Hawking--York counter-terms. While this can be seen more as a mathematical consequence of Chern--Simmons theorem, it physical significance might deserve a deeper analysis.}

At this stage, the boundary terms no longer act exclusively as elements necessary to make the variational problem well-posed, but are given the chance to shine a light on the hidden features of the actual theories.

Consider for instance an $f (R)$-theory, for which the action reads
\begin{equation}
\mathcal{S}_{\text{grav}} = \frac{1}{2\kappa} \int_{\mathcal{M}} f \left( R \right) \, \sqrt{-g} \, \mathrm{d}^4 x \;. \label{eq.fRaction}
\end{equation}
Assume as well, for sake of simplicity, that it is $f (R) = R^2$, or any other polynomial in the scalar curvature. It is possible to show that this theory requires at least a Gibbons--Hawking--York-like surface term of the form
\begin{equation}
\mathcal{S}_{\text{surf}} = \frac{1}{\kappa} \oint_{\partial \mathcal{M}} f'' \left( R \right) K \, \sqrt{\gamma} \, \mathrm{d}^3 x \;. \label{eq.fRboundterm}
\end{equation}

When the sum of Eqs.~\eqref{eq.fRaction} and~\eqref{eq.fRboundterm} is varied with respect to $g_{ab}$, cancellations similar to the Einstein--Hilbert case occur, and this is desirable, but eventually one is left with a term proportional to
\begin{equation}
f' \left( R \right) \delta R \;, \label{eq.varfRnoneras}
\end{equation}
which cannot be compensated by anything else, neither in the bulk action, nor in any further boundary term conceivable. If, then, one wants to recover the field equations, the only choice is to set $\delta R = 0$ on the boundary $\partial \mathcal{M}$, together with $\delta g_{\mu\nu}$.

What we are imposing here is, strictly speaking, the vanishing on the boundary of the variation of all the \emph{actual} degrees of freedom of the theory, in a scheme with second-order field equations only, as if we were still in the GR-case. But then, something is flawed with our initial formulation of the model, and the symbol $R$ in the action above signals nothing but the presence of another degree of freedom (at the bare minimum, a scalar field), ``hidden'' somewhere in the free components of $g_{\mu\nu}$.

This turns out to be the case, in fact, at least for all the $f (R)$-theories, which can almost everywhere be remapped into scalar-tensor models, with $f' (R)$ playing the role of the field $\phi$ in a Brans--Dicke theory~\cite{DeFelice:2010aj,Sotiriou:2008rp}.

This example might seem to point at the conclusion that the search for the proper form of the boundary terms can give precious hints about possible reformulations of some higher-order ETG's in terms of other, dynamically equivalent second-order theories, with manifest d.o.f.'s besides the metric (the latter carrying only the usual two).

Unfortunately, the result holding for $f(R)$-theories is almost unique, in the sense that the vanishing of uncompensated terms in the boundary terms does not lead, in general, to any further immediate identification of the additional d.o.f.'s, nor it allows for any easy identification of the geometric nature of the supplementary dynamical variables.

So, we ought not to overestimate the relevance of the diagnostic power of this ``tool''. Admittedly, it is true that, by looking at the variations of the boundary terms, it is possible to notice some telltales that a theory under examination is not as ``purely metric'' as promised by its action functional. Yet, as we shall soon see, there are much more powerful and fruitful techniques allowing to identify the precise nature of possible additional gravitational d.o.f.'s.

The gist here is that taking care of the boundary terms in an ETG is a necessary, preliminary step, which results in a well-posed variational formulation of the model. In a few cases (a very tiny subset, in fact), by simply looking at the boundary terms, it is possible to notice that something is hidden beneath a seemingly ``purely metric'' formulation, and the theory might be recast in terms of additional, non-metric (in the sense of ``non-gravitonic'') degrees of freedom. At the same time, as the complexity of the starting actions grows, it makes less and less sense to rely on the boundary term analysis to thoroughly grasp the ``true'' nature of the ETG itself.%
\section{The propagator for higher-curvature ETG's: linearization techniques\label{propagators}}

The first technique we review is based on the computation of the propagator for a generic higher-curvature ETG. To do this, we first perform a splitting of the metric tensor in the sum of a background metric $g^{(0)}_{\mu\nu}$, and a perturbation $\varepsilon \, h_{\mu\nu}$. Following a well-established tradition, the background configuration is assumed to be a maximally symmetric (MS) spacetime --- initially, a Riemann-flat one.

This method for extracting the number and type of gravitational d.o.f.'s benefits from a vast and comprehensive literature --- for recent contributions, see e.g. Refs.~\cite{Biswas:2014tua,Accioly:2002tz,buchbinder1992effective,Bartoli:1998gs} --- but its roots date back to seminal studies on quadratic corrections to the Einstein--Hilbert action~\cite{Stelle:1976gc,Stelle:1977ry}.
  
Herewith, we begin by showing how, starting from a generic higher-curvature theory, the contributions at order $\varepsilon^2$ (needed to compute the propagator), contain at most quadratic invariants in the Riemann tensor and its derivatives~\cite{Biswas:2014tua}. Then, we discuss the issue of gauge invariance when inverting the kinetic term for the metric tensor, obtaining an explicit form for the propagator. In the last part, we briefly explore whether the results thus collected can be extended to non-flat backgrounds, and make some additional comments.

\subsection{Linearization and quadratic gravity}

As we want to investigate higher-curvature, ``purely metric'' ETG's enforcing invariance under general coordinate transformations, the action must be a scalar function of the Riemann tensor and its covariant derivatives, albeit the form of the function can be quite general.


The metric $g_{\mu\nu}$ is then first split into the sum of its background value, plus a fluctuating/perturbation term, i.e.
\begin{equation}
g_{\mu\nu}=g^{(0)}_{\mu\nu} + \varepsilon h_{\mu\nu}.
\end{equation}
We take $g^{(0)}_{\mu\nu}$ to be the Minkowski metric $\eta_{\mu\nu}$. We are interested in the quadratic contributions (order $\varepsilon^2$), for they are the only terms contributing to the computation of the propagator. With this in mind, it can be shown that one does not need to consider the most general action containing the metric tensor, the Riemann tensor and its covariant derivatives given by Eq.~\eqref{genaction}, but it is enough to examine the following expression (see~\citep{Biswas:2014tua})
\begin{equation}\label{quadrF}
\begin{split}
\mathcal{S} = \frac{1}{\kappa} \int_\mathcal{M} &\sqrt{-g}\,\mathrm{d}^4 x\, \Big(\frac{R}{2} + \\
&+ R F_1(\square)R+R_{\mu\nu}F_2(\square)R^{\mu\nu}+R_{\mu\nu\rho\sigma}F_3(\square)R^{\mu\nu\rho\sigma}\Big).
\end{split}
\end{equation}

The content of Eq.~\eqref{quadrF} is nothing but a generalization of the theory advanced in Refs.~\cite{Stelle:1976gc,Stelle:1977ry}, where the coefficients of the higher-curvature terms are functions of the d'Alembertian operator, and we do not discard the term quadratic in the Riemann tensor.\footnote{Given the presence of the $F_3(\square)$-operator, the Gauss--Bonnet combination $\mathcal{G}$ cannot be deployed to express the quadratic term in the Riemann tensor as a combination of the other two quadratic invariants, $R^2$ and $R^{\mu\nu} R_{\mu\nu}$.}
Notice that, since the two Riemann tensors together are already of order $\varepsilon^2$ (${R^{(0)\alpha}}_{\beta\gamma\delta} = 0$), the covariant derivatives in the differential operator are just partial derivatives at the same order $\varepsilon^2$. This simplifies the action in Eq.~\eqref{quadrF} substantially, and the equations of motion for the perturbation field become, at that order,
\begin{equation}\label{eom}
\begin{split}
&a(\square)\square h_{\mu\nu}+2b(\square)\partial_\sigma\partial_{(\mu} h^\sigma_{\nu)}+c(\square)(\eta_{\mu\nu}\partial_\rho\partial_\sigma h^{\rho\sigma}+\partial_\mu\partial_\nu h)+\\
&+d(\square)\eta_{\mu\nu} h+f(\square)\square^{-1}\partial_\sigma\partial_\rho\partial_\mu\partial_\nu h^{\rho\sigma}=-2\kappa \tau_{\mu\nu},
\end{split}
\end{equation}
where $\tau_{\mu\nu}$ is the stress-energy-momentum tensor for matter (if it is present at all), and we have introduced the new symbols
\begin{subequations}\label{abcdf}
\begin{align}
&a(\square)=1+2F_2(\square)\square+8F_3(\square)\square,\\
&b(\square)=-1-2F_2(\square)\square-8F_3(\square)\square,\\
&c(\square)=1-8F_1(\square)\square-2F_2(\square)\square,\\
&d(\square)=-1+8F_1(\square)\square+2F_2(\square)\square,\\
&f(\square)=8F_1(\square)\square+4F_2(\square)\square+8F_3(\square)\square \,.
\end{align}
\end{subequations}
As pointed out in~\cite{Biswas:2014tua}, these functions are a generalized version of the coefficients found in~\cite{VanNieuwenhuizen:1973fi}. The functions $F_i(\square)$ must be analytic in the infrared limit in order to recover the GR regime. In particular, one requires that the conditions $\displaystyle\lim_{k^2\rightarrow 0} F_i(-k^2)\propto -k^2$ and $a(0)=-b(0)=c(0)=-d(0)=1$ , $f(0)=0$ hold.

\subsection{The propagator, and gauge fixings\label{propag}}

To compute the propagator in momentum space we have to invert the kinetic operator in Eq.~\eqref{eom}. This can be done by introducing a complete set of projectors $\{P^2,P^1,P^0_s,P^0_w\}$
for any symmetric rank-$2$ tensor, given by\footnote{For the sake of simplicity we will omit the tensorial structure where it is not needed.}
\begin{subequations}
\begin{align}
P^2&=\frac{1}{2}(\theta_{\mu\rho}\theta_{\nu\sigma})-\frac{1}{3}\theta_{\mu\nu}\theta_{\rho\sigma},\\
P^1&=\frac{1}{2}(\theta_{\mu\rho}w_{\nu\sigma}+\theta_{\mu\sigma}w_{\nu\rho}+\theta_{\nu\rho}w_{\mu\sigma}+\theta_{\nu\sigma}w_{\mu\rho}),\\
P^0_s&=\frac{1}{3}\theta_{\mu\nu}\theta_{\rho\sigma},\\
P^0_w&=w_{\mu\nu}w_{\rho\sigma},
\end{align}
\end{subequations}
where $\theta_{\mu\nu}$ and $w_{\mu\nu}$ are the transverse and longitudinal projectors in the momentum space, namely
\begin{equation}
\theta_{\mu\nu}=\eta_{\mu\nu}-\frac{k_\mu k_\nu}{k^2},\qquad w_{\mu\nu}=\frac{k_\mu k_\nu}{k^2}.
\end{equation}
To this set of operators we need to add the two ``transfer operators'' mapping quantities between spaces with the same spin --- see~\cite{Rivers1964}
\begin{equation}
P^0_{sw}=\frac{1}{\sqrt{3}}\theta_{\mu\nu}w_{\rho\sigma},\qquad P^0_{ws}=\frac{1}{\sqrt{3}}w_{\mu\nu}\theta_{\rho\sigma}.
\end{equation}

Every operator in Eq.~\eqref{eom} can then be expressed using the projectors $P_i$ by means of the combination $O=\sum\limits_{i=1}^6 c_i P^i$, and the equations of motion can be fully projected and rewritten as
\begin{equation}\label{eomproj}
\sum\limits_{i=1}^6 c_i P^i h_{\mu\nu} = \kappa (P^2+P^1+P^0_s+P^0_w) \tau_{\mu\nu}.
\end{equation}

Once we have the explicit form of the coefficients in Eq.~\eqref{eomproj}, we can use again each projector operator on the equations of motion; the orthogonality among the $P_i$'s allows then to get the final form of the propagator. From Eq.~\eqref{eom}, we reach the relations
\begin{subequations}
\begin{align}
&a k^2 P^2 h=\kappa P^2\tau \Rightarrow P^2 h=\kappa\left(\frac{P^2}{ak^2}\right)\tau,\label{spin2}\\
&(a+b)k^2 P^1 h=\kappa P^1\tau,\label{spin1}\\
&(a+3d)k^2 P^0_s h+(c+d)k^2\sqrt{3}P^0_{sw}h=\kappa P^0_s\tau,\label{spin0s}\\
&(a+2b+2c+d+f)k^2 P^0_w h+(c+d)k^2\sqrt{3}P^0_{ws}h=\kappa P^0_w\tau,\label{spin0w}
\end{align}
\end{subequations}
where $a,b,c,d,f$ are now to be considered functions of $k^2$, as we moved into momentum space.

For the spin-$2$ part, the propagator is found immediately (provided that $a\neq 0$), and it reads
\begin{equation}\label{spin2prop}
\Pi^{(2)}=\frac{P^2}{a k^2}.
\end{equation}

The propagators for the other components are less straightforward to determine. From the functions defined in Eq.~\eqref{abcdf}, we can see that some of the coefficients in front of the left-hand sides of Eqs.~\eqref{spin1}, \eqref{spin0s}, and~\eqref{spin0w} vanish identically. This is so because we are dealing with a gauge theory, and can be seen by imposing the Bianchi identities on the left-hand side of Eq.~\eqref{eom}. The calculation gives~\cite{Biswas:2014tua}
\begin{equation}\label{bianchi}
(a+b)\square h^\mu_{\;\nu,\mu}+(c+d)\square \partial_\nu h+(b+c+f)\square h^{\alpha\beta}_{\;,\alpha\beta\nu}=0,
\end{equation}
where the right-hand side is zero because of the conservation of $\tau_{\mu\nu}$. As it can be seen directly from Eq.~\eqref{abcdf}, the coefficients in front of each term are zero.

Now, the Bianchi identities are a byproduct of diffeomorphism invariance, and this implies in turn that the left-hand sides of Eq.~\eqref{spin1}, Eq.~\eqref{spin0w} and the mixing term in  Eq.~\eqref{spin0s} are singular, therefore  Eq.~\eqref{spin1} and Eq.~\eqref{spin0w} cannot be inverted directly. Nevertheless, this can be done for the spin-$2$ part and the spin-$0s$ part as it can be seen from Eqs.~\eqref{spin2},~\eqref{spin0s}; in particular, the latter reads
\begin{equation}\label{spin0sprop}
\Pi^{(0s)}=\frac{P^0_s}{(a-3c)k^2},
\end{equation}
where we have used the fact that $d=-c$.
Therefore, in the sub-space of the tensor product $(2\otimes 0s)$, the propagator is
\begin{equation}\label{prop}
\Pi=\frac{P^2}{a k^2}+\frac{P^0_s}{(a-3c)k^2}.
\end{equation}

Eq.~\eqref{prop} can be further rewritten as the sum of the standard GR propagator, plus additional terms. The GR propagator alone is given by
\begin{equation}
\Pi_{GR}=\frac{P^2}{k^2}-\frac{P^0_s}{2 k^2},
\end{equation} 
where the scalar component cancels out the longitudinal components of the graviton propagator. This part of the higher-order propagator is gauge independent, whereas to extract the other parts a gauge-fixing term is needed.\footnote{For an accurate treatment of the propagator of quadratic gravity (plus the standard Einstein--Hilbert term) on a flat background including the gauge fixing terms, see~\citep{Accioly:2002tz}. Also, as pointed out in~\cite{buchbinder1992effective}, the explicit form of the propagator depends in general on the definition of the fluctuating term $\varepsilon\,h_{\mu\nu}$.}

The complete propagator for a higher-curvature ETG can then be written as
\begin{equation}\label{propbis}
\Pi=\Pi_{GR}+\frac{1-a(-k^2)}{a(-k^2) k^2}P^2+\frac{1+\frac{a(-k^2)-3c(-k^2)}{2}}{[a(-k^2)-3c(-k^2)]k^2}P^0_s.
\end{equation}
Recalling that $a(0)=c(0)=1$, one has again that the infrared limit of the above formula corresponds to the bare GR-case, i.e. that $\displaystyle\lim_{k^2\rightarrow 0} \Pi\rightarrow\Pi_{GR}$. Therefore, the gauge-invariant part of the propagator of a generic higher-curvature ETG on a flat background contains the usual massless spin-$2$ part (the graviton), plus a certain number of additional degrees of freedom given by the zeros of the functions $a(-k^2)$ and $c(-k^2)$. Indeed, gauge invariance guarantees that it is
\begin{equation}
a(\square)=-b(\square),\;c(\square)=-d(\square),\;f(\square)=a(\square)-c(\square),
\end{equation}
and therefore just two arbitrary functions survive, to host the gravitational d.o.f.'s.

Looking at Eq.~\eqref{propbis}, we can draw some general conditions to constrain the additional propagating d.o.f.'s the theory might have. In particular, if $a(-k^2)-1=0$ (that is, if $F_2+4F_3=0$), there will be no additional propagating spin-2 particle other than the graviton. In the same way, if $1+\frac{a-3c}{2}=0$ (or equivalently, $3F_1+F_2+F_3=0$), there will be no additional scalar d.o.f.'s --- as we will see, these are sufficient but not necessary conditions.

To translate back the language of the propagators in the context of field theories, we provide a basic list of archetypal ETG's, organised by means of their propagator content. On some of the examples we shall elaborate again elsewhere in the paper.
\begin{itemize}
\item
$F_1(-k^2)=\alpha,\,F_2(-k^2)=F_3(-k^2)=0$, with $\alpha$ a constant. This is the case of $f(R)$-theories --- see Eq.~\eqref{quadrF}). We have that $a(-k^2)=1$ and $c(-k^2)=1+8\alpha k^2$, therefore the propagator is
\begin{equation}
\Pi=\Pi_{GR}+\frac{1}{2} \frac{P^0_s}{k^2+1/12\alpha}.
\end{equation}
A new scalar d.o.f. emerges, and it has non-tachyonic character (i.e., the square of the mass is positive) as long as $\alpha>0$.

\item
$F_i=\text{const}\neq 0$ and $F_1=F_3=-F_2/4$. The resulting action is proportional to the Gauss--Bonnet combination $\mathcal{G}$. It is $a(-k^2)=1$ and $c(-k^2)=1$, which ensures that such ETG has no additional degrees of freedom, and its propagator is the same as that of GR.

\item
$F_1(-k^2)=\alpha,\,F_2(-k^2)=\beta,\,F_3(-k^2)=\gamma$. This is the theory examined in Refs.~\cite{Stelle:1976gc,Stelle:1977ry} and it corresponds to an ETG with the most general correction up to quadratic curvature invariants without explicit dependence of differential operators. The square of the Riemann tensor can be traded for $R^2$ and the Ricci tensor squared after introducing the Gauss--Bonnet combination and a redefinition of the coefficients $\alpha,\beta$. The propagator becomes
\begin{equation}
\Pi=\Pi_{GR}-\frac{P^2}{k^2-m_0^2}+\frac{P_0^s}{2[k^2+m_2^2]},
\end{equation}
where $m_0^2=(2\beta+8\gamma)^{-1}$ and $m_2^2=(4\beta+12\alpha+4\gamma)^{-1}$. The propagator thus exhibits a new scalar term and a second spin-$2$ state. However, even if we fix the coefficients in such a way that the mass of the massive spin-$2$ is positive, the propagator will anyway have an overall minus sign, which is the telltale of the presence of a \emph{ghost} state --- i.e., a state with negative energy, see~\cite{Sbisa:2014pzo}, and the discussion in Sect.~\ref{ghosts}.

\item
$a(-k^2)=1-(k^2/m^2)^2,\,c=1-\frac{1}{3}(k^2/m^2)$, whence $8F_1(-k^2)+2F_2(-k^2)=-\frac{1}{3m^2}$ and $F_2(-k^2)+8F_3(-k^2)=\frac{1}{m^2}$. With this choice, one obtains the Einstein--Hilbert action plus a term proportional to the Weyl tensor squared, $C^{\alpha\beta\gamma\delta} C_{\alpha\beta\gamma\delta}$. The propagator is
\begin{equation}
\Pi=\Pi_{GR}-\frac{P^2}{k^2+m^2},
\end{equation}
and once again the propagator of the massive spin-$2$ comes with an overall minus sign, therefore it is a ghost state.

\item
$a(-k^2)=c(-k^2)$. For this particular choice, that corresponds to the condition $2 (F_1(-k^2)+F_3(-k^2))+F_2(-k^2)=0$, the propagator becomes
\begin{equation}
\Pi=\frac{1}{a(-k^2)}\Pi_{GR}.
\end{equation}
As long as the function $a(-k^2)$ has no zeros, the propagator does not develop any additional pole, i.e., the theory does not have additional states with respect to GR. Nevertheless, the function can be such that the ultraviolet behavior (large $k^2$ values) of the propagator is improved, e.g., if $a(-k^2)$ is a non-local entire function~\cite{Biswas:2014tua}.
\end{itemize}

Since the parameters $F_1,\,F_2$ and $F_3$ are functions of the d'Alembertian operator, one can consider, in addition to the previous examples, other models with improved ultraviolet behavior without the issue of ghost states, such as non-local theories (we shall come back on this point in Sect.~\ref{hamiltonian}).

One might ask whether the results just presented are valid also when studying perturbations around dS/AdS spacetimes, specifically when one talks about the emergence of ghost states. This is in general the case: if a ghost propagates on flat spacetime, then it can be considered as a feature of the full theory, at the non-linear level. The contrary is, unfortunately, not true. If there are no ghost states for a particular ETG on flat spacetime, this does not guarantee that they will not crop up on some other type of background metric.\footnote{See \cite{Nunez:2004ts} for a discussion about the presence of ghost states and light scalars in higher-curvature ETG's of the kind given by a Lagrangian density of the form $f(R,R_{\mu\nu}R^{\mu\nu},R_{\mu\nu\rho\sigma}R^{\mu\nu\rho\sigma})$ when linearizing around any MS spacetime.} In this case, a full non-linear analysis is needed to check once and for all whether the theory has ghosts or not. This issue will be considered again in the next Sections.

\subsection{Considerations on the method \label{consprop}}

After analyzing the theory of the propagator of a generic higher-curvature ETG at order $\varepsilon^2$ in the field expansion, we have concluded, following~\cite{Biswas:2014tua}, that such a modification of GR will in general present a massive, ghost-like spin-$2$ state (sometimes known as the Weyl \emph{poltergeist}), and an additional scalar d.o.f. that can also be a ghost~\cite{Nunez:2004ts}.

These conclusions agree with the fact that, when the functions $F_i(\square)$ are constant, the theory is equivalent to the one advanced in~\cite{Stelle:1976gc,Stelle:1977ry}. On the other hand, if the functions depend on $k^2$, then the theory can exhibit a richer structure. In particular, there can be more (or fewer) d.o.f.'s depending on the zeros of the functions $a(-k^2)$ and $c(-k^2)$, and in general some of such d.o.f.'s can again be ghost states.

Computing the propagator around flat space-time is a powerful and quite simple tool to identify the propagating degrees of freedom of a higher-curvature ETG, but it has to be used carefully. There is the possibility that, when considering the quadratic expansion~\eqref{quadrF}, some features of the fully non-linear model are lost. For instance, some pieces of the original action might have a vanishing quadratic term in the flat limit --- one of the simplest cases being $F(R)=R+\chi R^3$, see~\cite{Hindawi:1995cu} --- and hence disappear at the leading order. From other extraction methods (see the next Section) we already know that this particular model propagates the two helicity states of the graviton plus a scalar field. It turns out that, in the flat limit, the mass of the scalar field becomes infinite, and thus the corresponding propagator vanishes. Therefore, the study of the linearized theory is in general not sufficient to establish unambiguously which are the propagating degrees of freedom. Only a full non-linear analysis can answer the question in a definite way.%
\section{Auxiliary fields method\label{Auxiliary fields}}

We now turn to a different technique to count and extract the d.o.f.'s in a higher-order ETG whose (diff-invariant) action can depend on the Riemann tensor and its covariant derivatives. We focus on the so-called \emph{auxiliary fields method}, allowing to recast the given theory in a dynamically equivalent form made up by a standard GR-term, plus other non-metric variables, all yielding second-order only field equations.

This method has been widely used in many contexts~\cite{Deruelle:2009zk,Hindawi:1995an,Hindawi:1995cu,Balcerzak:2008bg,Chiba:2005nz,Baykal:2013gfa,Rodrigues:2011zi}. In principle, it is a powerful technique not requiring any linearization to extract information about the theory at hand. 
We shall see, however, that once the theory has been recast into its second-order form the question remains open, of what kind of d.o.f.'s are encoded in the auxiliary fields.

The most general diffeomorphism-invariant action for the metric tensor $g_{\mu\nu}$ is given by Eq. \eqref{genaction}, that we rewrite below for convenience 
\begin{equation}
\mathcal{S} = \frac{1}{2 \kappa}\int_\mathcal{M} \sqrt{-g}\,\mathrm{d}^4 x \, f\left(g_{\mu\nu},R_{\mu\nu\rho\sigma},\nabla_{\alpha_1} R_{\mu\nu\rho\sigma},...,\nabla_{(\alpha_1}...\nabla_{\alpha_m)}R_{\mu\nu\rho\sigma}\right).
\end{equation} 
The metric variation of the this action yields in general higher-order equations of motion for $g_{\mu\nu}$, which is a drastic departure from the typical dynamical models for particles and fields, based on second-order evolution equations. It might then be desirable to translate the higher-order dynamics of such ETG's in a more traditional setting, possibly redistributing the d.o.f.'s into a new set of variables.

In what follows, we show that, while the auxiliary fields method can sometimes help with the issues with Eq.~\eqref{genaction}, in most of the cases the reformulated action will still be too complicated, making it necessary to resort to other tools to ultimately extract the sought-after d.o.f.'s. 

We begin by testing the method in the context of fourth-order ETG's, then we elaborate on how the technique can be applied also to higher-order theories.

\subsection{Fourth-order gravity}

To begin with, we specialize Eq.~\eqref{genaction} so the sub-class of higher-curvature ETG's given by
\begin{equation}\label{action}
\mathcal{S} = \frac{1}{2\kappa}\int_\mathcal{M} \sqrt{-g}\,\mathrm{d}^4 x \, f(g_{\alpha\beta} , R_{\mu\nu\rho\sigma}).
\end{equation}
The equations of motion read (bracketed indices denote symmetrisation with respect to the enclosed pair, regardless of their contra-/co-variant position)
\begin{equation}\label{4ordeom}
R^{(\mu}_{\;\;\;\alpha\rho\sigma}\frac{\partial f}{\partial R_{\nu)\alpha\rho\sigma}}-2\nabla_\rho\nabla_\sigma\frac{\partial f}{\partial R_{\rho(\mu\nu)\sigma}}-\frac{1}{2}fg^{\mu\nu}=0 \,,
\end{equation}
and they contain fourth-order derivatives of the metric tensor.

The auxiliary-fields method aims at rewriting the action above in the form
\begin{equation}\label{actionaux}
\mathcal{S} = \frac{1}{2\kappa}\int_\mathcal{M} \sqrt{-g}\,\mathrm{d}^4 x \,\left[f\left(\rho_{\mu\nu\rho\sigma}\right)+\frac{\partial f}{\partial\rho_{\mu\nu\rho\sigma}}\left(R_{\mu\nu\rho\sigma}-\rho_{\mu\nu\rho\sigma}\right)\right] \,,
\end{equation}
where the field $\rho_{\mu\nu\rho\sigma}$ is considered independent of the metric tensor, and has all the symmetries of the Riemann tensor. To the action for the gravitational sector, one adds the action for matter fields, which is assumed to depend only on $g_{\mu\nu}$, hence $\mathcal{S}_{\text{matter}} = \mathcal{S}_{\text{matter}} \left[ g_{\mu\nu} \right]$.

Variations of Eq.~\eqref{actionaux} with respect to $g_{\mu\nu}$ and $\rho_{\mu\nu\rho\sigma}$ yields the following equations of motion (for the moment, we ignore all the issues related to the presence, or lack, of the boundary terms), where also the variation of the matter sector has been performed,
\begin{subequations}
\begin{align}
&E_{\mu\nu}=T_{\mu\nu},\label{genE}\\
&\frac{\partial^2 f}{\partial\rho_{\mu\nu\rho\sigma}\partial\rho_{\alpha\beta\gamma\delta}}\left(R_{\alpha\beta\gamma\delta}-\rho_{\alpha\beta\gamma\delta}\right)=0 \,.\label{aux}
\end{align}
\end{subequations}
In Eq.~\eqref{genE} above, $E_{\mu\nu}$ is a ``generalized Einstein tensor'', and $T^{\mu\nu}$ is the stress-energy-momentum tensor of matter. The equivalence between~\eqref{action} and~\eqref{actionaux} is true everywhere (\emph{on-shell}) except for the values of the field $\rho_{\mu\nu\rho\sigma}$ for which the second derivative in~\eqref{aux} is zero. Those ``points'' (in fact, field configurations) will generate a certain number of (inequivalent) subsets in which the rewriting in terms of auxiliary fields is valid.\footnote{As pointed out in~\cite{Deruelle:2009zk}, one could also consider the action as a function of $R_{\mu\nu\rho\sigma}$ and of \emph{two} auxiliary fields, $\rho_{\mu\nu\rho\sigma}$ and $\phi_{\mu\nu\rho\sigma}=\frac{\partial f}{\partial\rho_{\mu\nu\rho\sigma}}$, in such a way that one is able to treat all the sub-cases in a unified manner.}

Typically, in order to put the action~\eqref{actionaux} in a canonical form --- i.e., with canonical kinetic terms for the auxiliary fields --- at least one additional step is required. Indeed, one must perform a field redefinition, and introduce a suitably reformulated metric tensor; see the next paragraphs and Refs.~\cite{Hindawi:1995an,Hindawi:1995cu}. A few specific, well-known examples~\cite{Sotiriou:2008rp,Hindawi:1995an,Hindawi:1995cu} will help clarifying this point.

\subsubsection{$f(R)$-theories}

The simplest and most common case is given by $f(R)$-theories --- see~\cite{Sotiriou:2008rp} for a comprehensive review. Such models are computationally manageable, but they have a structure rich enough to explore all the features of the method we are discussing. In 4 spacetime dimensions, the action reads
\begin{equation}\label{fR}
\mathcal{S}=\frac{1}{2 \kappa}\int_\mathcal{M} \sqrt{-g}\,\mathrm{d}^4 x \,f(R).
\end{equation}
Upon introducing an auxiliary field, the previous line can be rewritten as
\begin{equation}\label{fRaux}
\mathcal{S}=\frac{1}{2\kappa}\int_\mathcal{M} \sqrt{-g}\,\mathrm{d}^4 x\,\left[f(\psi)+f'(\psi)(R-\psi)\right],
\end{equation}
where the prime denotes a derivative with respect to $\psi$.

Variation with respect to the auxiliary field results in
\begin{equation}
f''(\psi)(R-\psi)=0.
\end{equation}
Therefore, the equivalence between Eqs.~\eqref{fR} and~\eqref{fRaux} is ensured on-shell, except for those values of $\psi$ for which $f''(\psi)=0$ --- this is a sufficient but not necessary condition~\cite{Sotiriou:2008rp}. Intervals between these ``points'' define different \emph{sectors} of the theory, i.e. inequivalent scalar-tensor representations of the very same dynamical content and behaviour (not to be confused with gravitational/matter sectors). By introducing a new variable defined as $\phi=f'(\psi)$, the action takes the form
\begin{equation}\label{BranseD}
\mathcal{S} = \frac{1}{2\kappa}\int_\mathcal{M} \sqrt{-g}\,\mathrm{d}^4 x\,\left[\phi R-V(\phi)\right],
\end{equation}
where 
\begin{equation}
V(\phi)=\psi(\phi)\phi-f(\psi(\phi)).
\end{equation}
In order for this transformation to be invertible, we require once again that $f''(R)\neq 0$. Now the theory has the precise aspect of a scalar-tensor ETG of the Brans--Dicke type with $\omega=0$. Hence, the seemingly ``purely metric'' action~\eqref{fR} has been translated into a dynamically equivalent model containing the standard GR-contribution (massless spin-$2$ graviton), plus an additional scalar d.o.f.  The equations of motion obtained from Eq.~\eqref{BranseD} read
\begin{subequations}
\begin{align}
&G_{\mu\nu}=\frac{1}{\phi}\left[\nabla_\mu \nabla_\nu \phi-g_{\mu\nu}\left(\square \phi -V(\phi)/2\right)\right],\\
&3 \square \phi+2V(\phi)-\phi\frac{d V}{d \phi}=0 \,,
\end{align}
\end{subequations}
where $G_{\mu\nu}$ is the standard Einstein tensor. Hence, this ETG has been effectively reduced to a theory with only second-order equations of motion; the dynamical content is the same (three d.o.f.'s per each model), but the representation has shifted from a higher-derivative arrangement, to a non-minimally coupled second-order structure.

For each \textit{sector} of the newly-obtained scalar-tensor theory, one can perform a conformal transformation on the metric and a new $\phi$-field redefinition to provide a canonical kinetic term for the scalar part. Such manipulations are given by
\begin{subequations}\label{conftrans}
\begin{align}
&\tilde{g}_{\mu\nu}\equiv\phi \, g_{\mu\nu},\\
&\tilde{\phi}\equiv \sqrt{\frac{3}{2\kappa}} \log\phi.
\end{align}
\end{subequations}
The action then becomes
\begin{equation}
S=\int_\mathcal{M} \sqrt{-\tilde{g}}\, \mathrm{d}^4 x\, \left[\frac{\tilde{R}}{2\kappa}-\frac{1}{2}\partial_\alpha \tilde{\phi} \, \partial^\alpha \tilde{\phi}-U(\tilde{\phi})\right].
\end{equation}

Despite its simplicity, the case of $f(R)$-ETG's already allows to highlight one key critical point. To this end, let us consider the specific case where $f(R):=R+\alpha R^3$. The auxiliary field is then given by $\phi=1+\alpha R^2$, and the condition for the invertibility becomes $\alpha R\neq 0$. Therefore, the equivalence between the scalar-tensor theory and the original ETG is not guaranteed e.g. in Minkowski spacetime (where $R=0$ identically), which explains why the linearization method fails to identify the extra d.o.f. when applied around Minkowski spacetime (see Section~\ref{propagators}).

\subsubsection{Quadratic gravity \label{quadgrav}}

Another interesting example is offered by the case of quadratic corrections to Einstein GR~\cite{Stelle:1976gc}. The action is given by
\begin{equation}\label{R2}
\mathcal{S}=\frac{1}{2 \kappa}\int_\mathcal{M} \sqrt{-g}\,\mathrm{d}^4 x \,\left[R+\alpha R^2+\beta R_{\mu\nu} R^{\mu\nu}+\gamma R_{\mu\nu\rho\sigma}R^{\mu\nu\rho\sigma}\right].
\end{equation}
Using the definition and properties of the Weyl tensor, one can prove that
\begin{equation}
C^{\mu\nu\rho\sigma} C_{\mu\nu\rho\sigma} = R_{\mu\nu\rho\sigma} R^{\mu\nu\rho\sigma} - 4 R_{\mu\nu} R^{\mu\nu} + \frac{R^2}{3} \,,\label{weyldecomp}
\end{equation}
and by dropping a term which is proportional to the Gauss--Bonnet invariant $\mathcal{G}$,\footnote{Recall that, in $4$ spacetime dimensions, the Gauss--Bonnet combination $\mathcal{G}$ is a topological invariant, hence can be added and/or subtracted without affecting the resulting field equations --- naturally, apt boundary terms must be introduced as well~\cite{Padmanabhan:2010zzb}.} Eq.~\eqref{R2} can be rewritten as
\begin{equation}\label{R2bis}
\mathcal{S}=\frac{1}{2 \kappa}\int_\mathcal{M} \sqrt{-g}\,\mathrm{d}^4 x \,\left[R+\frac{1}{6 m_0^2}R^2-\frac{1}{2 m_2^2} C^{\mu\nu\rho\sigma} C_{\mu\nu\rho\sigma}\right],
\end{equation}
where $m_0^{-2}=6\alpha+2\beta+2\gamma$ and $m_2^{-2}=-\beta-4\gamma$. Following the procedure found in~\cite{Hindawi:1995an}, it is more convenient to study the two correction terms separately, while still keeping the standard Einstein--Hilbert term in the action.

The first correction is tantamount to an $f(R)$-theory, and therefore we know that it can be reformulated in terms of GR, plus an additional scalar field. \cite{Hindawi:1995an} proves that, for $m_0>0$, the theory has a stable minimum at a vanishing value of the (canonically normalized) scalar field, and $m_0$ is in fact the mass of the perturbations.

The correction given by the Weyl-squared term corresponds to an additional massive spin-$2$ field. Indeed, using Eq.~\eqref{weyldecomp}, the action~\eqref{R2bis} (without considering the $R^2$-term) can be rewritten as
\begin{equation}\label{weyl}
\begin{split}
\mathcal{S}&=\frac{1}{2 \kappa}\int_\mathcal{M} \sqrt{-g}\,\mathrm{d}^4 x \,\left[R-\frac{1}{2 m_2^2}C^{\mu\nu\rho\sigma} C_{\mu\nu\rho\sigma}\right]\\
&=\frac{1}{2 \kappa}\int_\mathcal{M} \sqrt{-g}\,\mathrm{d}^4 x \,\left[R-\frac{1}{m_2^2}\left(R_{\mu\nu}R^{\mu\nu}-\frac{1}{3}R^2\right)\right]\\
&=\frac{1}{2 \kappa}\int_\mathcal{M} \sqrt{-g}\,\mathrm{d}^4 x \,\left[R-G_{\mu\nu}\pi^{\mu\nu}+\frac{1}{4}m^2_2(\pi^{\mu\nu}\pi_{\mu\nu}-\pi^2)\right] \,,
\end{split}
\end{equation}
where the auxiliary field on-shell is given by 
\begin{equation}
\pi_{\mu\nu}=\frac{2}{m_2^2}\left(R_{\mu\nu}-\frac{1}{6}g_{\mu\nu}R\right) \,,
\end{equation}
and it satisfies a direct generalization to curved space-time of the Fierz--Pauli conditions~\cite{Fierz:1939ix}. The latter characterize completely a spin-$2$ field, and can be obtained via the formal substitutions~\cite{Hindawi:1995an}
\begin{equation}
\begin{cases}
\partial^\mu \phi_{\mu\nu}=0\\
\eta^{\mu\nu} \phi_{\mu\nu}=0 
\end{cases} \longrightarrow
\begin{cases}
\nabla^\mu \phi_{\mu\nu}=0\\
g^{\mu\nu} \phi_{\mu\nu}=0 \end{cases} \;.
\end{equation}

The auxiliary-fields approach also provides other relevant results. For instance, the ``$R$ plus Weyl-squared'' ETG can be recast in canonical form by generating kinetic terms for the non-metric auxiliary field, and reducing the curvature terms to the standard Einstein--Hilbert one. This is accomplished by generalizing the conformal transformation~\eqref{conftrans} used in the case of $f(R)$-theory~\cite{Hindawi:1995an}. In this way, it is possible to identify the mass of the spin-$2$ field, and show that the latter is always a ghost. It should be noticed that, if $\beta=4\gamma$, the Ricci and the Riemann tensors in \eqref{R2} can be eliminated completely using the Gauss--Bonnet combination and that, in this limit, the mass of the spin-$2$ field goes to infinity accordingly.

Such conclusions hold as well when one considers the general quadratic action~\eqref{R2bis}.\footnote{To show this, some additional manipulations are needed due to the fact that there are now couplings between the scalar and spin-$2$ kinetic energy terms. See~\cite{Hindawi:1995an} for the details.} Once again, it turns out that the massive spin-$2$ is a ghost, whereas the graviton and the scalar degrees of freedom are not.

Finally, following again an analysis similar to the $f(R)$-case, one can find the range of free parameters for which a second-order ETG where auxiliary fields are introduced, is equivalent to the original higher-curvature one.

\subsubsection{General functions of Ricci and Riemann tensors\label{genfricciriem}}

In principle, the auxiliary-fields method can be extended to more complicated ETG's. The counting of the d.o.f.'s can be estimated as follows~\cite{Hindawi:1995cu}: the symmetries and diff-invariance of $g_{\mu\nu}$ reduce to six the free components of the metric. Thus, when solving the Cauchy problem for a system of fourth-order field equations, six initial conditions are required. Using then the auxiliary fields as in the previous paragraphs, it is possible to prove~\cite{Hindawi:1995cu} that assigning the initial conditions for these fields is tantamount to fixing the second and third derivatives of $g_{\mu\nu}$ in the original representation. Therefore, the auxiliary fields can carry at most $6$ d.o.f.'s, whereas the metric carries the remaining $2$, as in standard GR. In this way we get a first, raw upper bound of eight d.o.f.'s for the class of theories described by Eq.~\eqref{action}. 

As an example, let us consider now a different restriction of the action~\eqref{action}, given by an arbitrary function of the Ricci tensor $f(R_{\mu\nu})$. One can introduce an auxiliary field given by a tensor $X_{\mu\nu}$ with the same symmetries of $R_{\mu\nu}$, as done in~\citep{Hindawi:1995an}. This leads to
\begin{equation}
\begin{split}
\mathcal{S}&=\frac{1}{2\kappa}\int_\mathcal{M} \sqrt{-g}\,\mathrm{d}^4 x \,f(R_{\mu\nu})=\\
&=\frac{1}{2\kappa}\int_\mathcal{M} \sqrt{-g}\,\mathrm{d}^4 x \,\left[f(X_{\mu\nu})+\frac{d f}{d X_{\mu\nu}}(R_{\mu\nu}-X_{\mu\nu})\right]=\\
&=\frac{1}{2\kappa}\int_\mathcal{M} \sqrt{-g}\,\mathrm{d}^4 x \,[f(X_{\mu\nu}(\pi_{\rho\sigma}))+\pi_{\mu\nu}(R_{\mu\nu}-X_{\mu\nu}(\pi_{\rho\sigma}))].
\end{split}
\end{equation}
We expect at most six additional d.o.f.'s besides those encoded in the massless graviton, to be found inside the tensor $\pi^{\mu\nu}=\frac{d f}{d X_{\mu\nu}}$ (or, equivalently, in $X_{\mu\nu}$). The introduction of the auxiliary fields requires the non-degeneracy condition 
\begin{equation}
\det \frac{d^2 f}{d X_{\mu\nu} d X_{\rho\sigma}}\neq 0,
\end{equation}
to hold true. This requirement will again generate different sectors of the theory, and in each of them one is supposed to define an appropriate auxiliary field $\pi_{\mu\nu}$.

It is not difficult to check that the tensor $X_{\mu\nu}$ carries indeed $6$ d.o.f.'s. By construction, it is a symmetric tensor, and therefore it contains no more than $10$ independent components. Then, since on-shell we have $X_{\mu\nu}=R_{\mu\nu}$, the Bianchi identity ensures that $\nabla^{\mu}(X_{\mu\nu}-\frac{1}{2}g_{\mu\nu}X)=0$, and these provide $4$ constraints. Hence, $X_{\mu\nu}$ has six independent components.

Unfortunately, for the case of a generic function $f$ it is not possible to further separate those d.o.f.'s according to some fixed recipe. To show that they can be rearranged into a massive spin-$2$ field plus a massive scalar field, one needs to linearize the theory around an appropriate maximally symmetric spacetime --- see~\cite{Hindawi:1995cu} and Sect.~\ref{propagators}. Similar considerations hold for a general action of the type~\eqref{action}.

Before moving on to the next section, let us briefly point out some potential issues related to a na{\"i}ve use of the auxiliary fields. For most of the instances considered in this section, the introduction of auxiliary fields allowed us to clearly identify the number and nature of the additional d.o.f.'s in a non-perturbative fashion. This might not be the case in general.

Let us consider, for example, the action \eqref{action} rewritten as a function of the curvature invariants in the following fashion
\begin{equation}\label{multisc}
\mathcal{S}=\frac{1}{2\kappa}\int_\mathcal{M} \sqrt{-g}\,\mathrm{d}^4 x \,f(X_i),
\end{equation}
where the $X_i, i =1,\dots,n$ are various scalars constructed out of the Riemann and the metric tensors. Following the usual protocol, it is possible to introduce a certain number of auxiliary scalar fields, to expose the presence of possible additional d.o.f.'s. In particular, one can write a dynamically equivalent action involving a certain number of scalar fields $\Phi_j$, non-minimally coupled to the curvature scalars $X_i$, plus the potentials for the $\Phi_j$'s. The action resulting from this manipulation can still be very complicated --- see~\citep{Chiba:2005nz} for a concrete example --- and in general one needs to perform additional manipulations to further simplify the model  and extract the number and nature of the d.o.f.'s (by linearizing, for instance). Moreover, it is not obvious that this na{\"i}ve way of introducing auxiliary fields actually reduces the order of the equations of motion for the metric tensor (which for \eqref{multisc} are of fourth order). Therefore, instead of exposing the presence of additional d.o.f.'s, the auxiliary fields only provide an alternative, potentially more convoluted description of the dynamics of the model.

\subsection{Beyond fourth-order gravity}


Eq.~\eqref{action} contains all possible curvature invariants constructed with the Riemann tensor and the metric tensor, but nonetheless these combinations will still produce equations of motion whose highest order is the fourth.\footnote{This is because adding more curvature tensors and their contractions does not alter the order of derivatives of $g_{\mu\nu}$ appearing in the action --- such order always remains equal to two.} To move forward with our analysis and consider more general theories, it is then necessary to take into account an explicit dependence of the action on differential operators acting on the Riemann tensor, as done, for instance, in Refs.~\cite{Wands:1993uu,Brown:1995su}.

In the next section we will discuss how to introduce auxiliary fields for a general diffeo-invariant action as~\eqref{genaction}, which includes differential operators acting on the Riemann tensor. We start by noticing that, in principle, a term of the form $\square R$ contains up to fourth derivatives of the metric tensor, hence it ought to have been included in the previous discussion. Yet, the term $\square R$ is in fact a covariant total divergence and, as we are not considering possible issues with boundary terms, we can safely ignore contributions of this type for the moment.

\subsubsection{General higher-order theories}

Starting from the very general premise of a diff-invariant ETG with action as in Eq.~\eqref{genaction}, it is possible~\cite{Brown:1995su} to significantly simplify the Lagrangian by introducing tensorial auxiliary variables, and then integrating by parts (while also discarding boundary terms). Successive iterations of such protocol lead to the remarkable disappearance of all the derivatives of the Riemann tensor; even the Riemann tensor itself can be made drop out. The final action reads
\begin{equation}\label{genscaltens}
\begin{split}
\mathcal{S}=\frac{1}{2 \kappa}\int_\mathcal{M} &\sqrt{-g}\,\mathrm{d}^4 x \,\left\lbrace\mathcal{L}(g_{\mu\nu},V^{(0)}_{\bullet},\dots,V^{(m)}_{\bullet}) + U^{(0)\,\bullet} \left(R_{\bullet} - V^{(0)}_{\bullet} \right)\right.\\
&\left.-\left[(\nabla_{\! \bullet}U^{(1)\,\bullet}) V^{(0)}_\bullet + U^{(1)\,\bullet} V^{(1)}_{\bullet} \right] - \dots\right. \\
&\left.- \left[(\nabla_{\! \bullet} U^{(m)\,\bullet}) V^{(m-1)}_{\bullet} + U^{(m)\,\bullet} V^{(m)}_{\bullet}\right]\right\rbrace,
\end{split}
\end{equation}
and it now depends only on the metric tensor and the auxiliary tensor fields $U^{(i)}$ and $V^{(i)}$, $i = 1,\dots,m$. In the previous formula, the bullet symbol ``$\bullet$'' stands for any combination of indices such that the contractions make sense and eventually create a scalar quantity.\footnote{From the first line of Eq.~\eqref{genscaltens} it results that $U^{(0)}$ and $V^{(0)}$ both are $4$-index quantities, as they contract with the Riemann tensor $R_{\alpha\beta\gamma\delta}$. Hence, $U^{(1)}$ and $V^{(1)}$ each sport $5$ indices, as their contraction demand one more index than $U^{(0)}$ and $V^{(0)}$, and so forth.}

We can calculate the equations of motion for the auxiliary variables and use them in the action~\eqref{genscaltens} to recover Eq.~\eqref{genaction}. This procedure is again very powerful in principle, but in most of the cases not very helpful. Apart from some simple yet relevant ETG's, such protocol does not help in building a general and effective recipe to isolate and identify the additional d.o.f.'s of the higher-curvature theory (and decide whether they are dynamical or not).

As an example, let us consider, with Refs.~\cite{Hindawi:1995cu,Wands:1993uu,Barth:1983hb,Schmidt:1990dh,Amendola:1993bg}, an ETG which involves derivatives of the curvature scalar, i.e. a $\square^k R$-model; this archetype will allow us to outline a few delicate points. The order of the field equations is determined here by the number $k$, which affects as well the number of additional auxiliary variables. Every new instance of the d'Alemebertian operator carries two more time derivatives $\partial_0^2$. Hence, we might na{\"i}vely expect one additional degree of freedom for each power of the box operator. On the other hand, we already know that every time a term in the action is a total divergence, it will not contribute to the equations of motion, as it occurs precisely with a pure $\square^k R$-term. A term of the kind $\square^k R\,\square^j R$, instead, yields an object of the form $R\,\square^{k+j} R$ (after integration by parts), and that actually contributes to the equations of motion.

Regarding these kinds of corrections, it has been shown, both at the level of the equations of motion~\cite{Wands:1993uu,Barth:1983hb,Schmidt:1990dh,Amendola:1993bg}, and of the action~\cite{Hindawi:1995cu}, that an ETG of the type
\begin{equation}
\mathcal{S}=\frac{1}{2\kappa}\int_\mathcal{M} \sqrt{-g}\,\mathrm{d}^4 x \,f\left(R,\square R,\square^2 R,...,\square^k R\right),
\end{equation}
can in general be rewritten as a theory describing a set of scalar fields non-minimally coupled to standard GR. The number of auxiliary non-metric d.o.f.'s can either be $2k+1$ or $2k$, based on the emerged functional dependencies in the translated action.\footnote{As usual, the introduction of Lagrange multipliers and auxiliary fields requires the fulfillment of non-degeneracy conditions to ensure the equivalence with the original higher-curvature model. This procedure generates different sectors as in the case of $f(R)$ theory~\cite{Hindawi:1995cu}.} Upon writing the function as $f(\lambda,\lambda_1,...,\lambda_{k+1})$, if $\partial f/ \partial(\lambda_{k+1})$ is a function of $\lambda_{k+1}$, then we are in the first case and $k$ scalar fields are ghost-like, whereas the remaining $k+1$ are not. If instead $\partial f/ \partial(\lambda_{k+1})$ is not a function of $\lambda_{k+1}$, then it is a function of $\lambda_k$, in which case we arrive at $2k$ new scalar fields, of which \textit{at least} $k-1$ are ghost-like.

With this premise, let us look at the theory for which $\mathcal{L}=\sqrt{-g}\left(R+\gamma R\square R\right)$. It falls into the first category outlined above, therefore we expect $2\cdot k=2$ additional scalar fields to be present in the theory. The reformulated action is given by~\cite{Wands:1993uu}
\begin{equation}
\mathcal{S}=\frac{1}{2\kappa}\int_\mathcal{M} \sqrt{-g}\,\mathrm{d}^4 x \,\left[\left(1+\gamma \phi_1+\gamma \square \phi_0\right)R-\gamma \phi_0\phi_1\right],
\end{equation}
ignoring possible couplings with the standard matter fields. The field equations for the scalar fields read
\begin{subequations}
\begin{align}
&\gamma \square R=\gamma \phi_1\,,\\
&\gamma R=\gamma \phi_0\,,
\end{align}
\end{subequations}
hence the non-degeneracy condition is $\gamma\neq 0$. Introducing a new scalar defined as $\Phi=(1+\gamma \phi_1+\gamma \square \phi_0)$, the action can be rewritten as
\begin{equation}
\mathcal{S}=\frac{1}{2\kappa}\int_\mathcal{M} \sqrt{-g}\,\mathrm{d}^4 x \,\left[\Phi R-\phi_0\left(\Phi-1\right)+\gamma \phi_0\square\phi_0\right].
\end{equation}
The expression above can be further manipulated to generate a canonical kinetic term for the scalar field, resulting in
\begin{equation}
\mathcal{S}=\int_\mathcal{M} \sqrt{-g}\,\mathrm{d}^4 x \,\left[\frac{\Phi R}{2\kappa}+\frac{1}{2} \psi\square\psi-\frac{1}{\sqrt{4\kappa\gamma}}\psi\left(\Phi-1\right)\right],
\end{equation}
where $\psi = \phi_0 \sqrt{\gamma/\kappa}$.
This is the action of a Brans--Dicke theory with $\omega=0$, plus an additional scalar field with an interaction potential. Notice the absence of ghost states, as the condition $k-1=0$ preventing the onset of instabilities is here a built-in feature.

In a similar manner, one can show that the theory whose Lagrangian reads $\mathcal{L}=\sqrt{-g}\left(R+\alpha R^2+\gamma R\square R\right)$ can be remapped into a scalar-tensor theory with two dynamical additional scalar fields~\cite{Gottlober:1989ww}. Before going to the conclusion of this section, it is worth noticing that the dynamical content of the last two examples can be also extracted linearizing the theory around the Minkowski background using the techniques reviewed in the previous Sect.~\ref{propagators}. In doing so, it is easy to show that the scalar part of the propagator \eqref{propbis} possesses two additional poles, hence in agreement with the results obtained at the non-linear level using the auxiliary fields method.

\subsection{Considerations on the method}

We have seen how, using the auxiliary-fields method, it is possible to recast many sub-classes of higher-curvature ETG's as second-order theories for the metric tensor (now sporting only the $2$ d.o.f.'s of standard GR), plus some more fields. In some cases, this technique allows us to identify not only the number, but also the nature of the additional d.o.f.'s coupled to the graviton --- e.g. in $f(R)$-theories, quadratic gravity, or even when covariant derivatives of the Riemann tensor are allowed in the action.

Very often, however, this method is not powerful enough to fully explore the features of a given higher-curvature ETG, and one has to rely on other tools (e.g. linearization) to bring the analysis to its end.

On the bright side, we have been able to conclude that fourth-order gravity generally admits $8$ d.o.f.'s at most, sub-divided into one massless spin-$2$ field (the graviton), one ghost-like massive spin-$2$ field (the mentioned Weyl poltergeist), and one scalar field. At the same time, we have seen (in Sect. \ref{quadgrav}) that for some special cases the poltergeist field (which carries $5$ d.o.f.'s) can be made disappear.

Moving past fourth-order theories, we have treated models involving terms of the $\square^k R$-type, recast as GR plus additional scalar fields --- a certain number of the latter behaving inevitably as ghosts. More complicated higher-order ETG's might still be studied with this method, at least in principle, but the level of complexity grows rapidly, and it gets harder and harder to make definite statements. Again, after finding the propagator of the linearized corresponding model, one can go a bit farther, but soon comes the moment when also the auxiliary-fields technique hits the wall, and cannot advance any further.
\section{Expansion around maximally symmetric spacetimes \label{linearization}}

Our discussion is progressing from less general extraction techniques (e.g. propagators, and hence linearizations), to more refined protocols (e.g. the auxiliary-fields method), in an attempt at framing the ultimate recipe to determine the correct number and representations for the gravitational d.o.f.'s of a given higher-order ETG. The improvement brought in by the auxiliary-fields method has been the possibility for a non-linear analysis of the dynamical content. Still, already when considering actions of the type
\be\label{fRiem}
\mathcal{S} = \frac{1}{2\kappa} \int_{\mathcal{M}} f(g_{\mu\nu},R_{\alpha\beta\gamma\delta}) \, \sqrt{-g}\, \mathrm{d}^4 x \;,
\ee
the introduction of the auxiliary fields falls short of our expectations, and the explicit emergence of the sought-after d.o.f.'s can only be achieved by applying two or more techniques separately.

One can further improve the situation by mixing together the best features of the two protocols discussed so far, namely the expansion of the action, and the apt reformulation of the Lagrangians in terms of curvature invariants. Such method makes it possible to deal more safely with ETG's of the type~\eqref{fRiem}.

The protocol we are mentioning --- see Refs.~~\cite{Hindawi:1995cu,Chiba:2005nz,Capozziello:2013ava} --- consists in expanding the action of a given ETG up to second order (in curvature invariants)  around a specific type of background solution, as long as such solution is admitted by the theory at the full non-linear level. In this case, the spacetime acting as the ``ground level''  must be a maximally symmetric (MS) solution, i.e. one characterised by a constant value of the scalar curvature $R$ (MS solutions include Minkowski spacetime as a sub-case, and also de Sitter and anti-de Sitter solutions)\footnote{The linearization of the equations of motion, around MS spacetimes, have also been used in the literature to classify higher-order theories of the type of Eq.\eqref{fRiem} on the basis of their spectrum, see Refs~\cite{Bueno:2016ypa,PhysRevD.93.101502}.}.

The outcome of the procedure is a quadratic ETG, its specific form depending on the choice of the initial action, for which it is easier to determine the dynamical content and its possible representations. In this sense, this ``non-linear expansion method'' deploys the power of the linearization and the generality of a fully non-linear tool.

Once again, the merits of the method must be welcomed with a pinch of salt: we shall thoroughly elaborate on the pitfalls of such technique, some of which are in plain sight, whereas some others remain hidden to a first inspection. Also, we shall compare the range of results one can expect from this recipe, with what can be achieved from the other procedures.

\subsection{A few milestones of quadratic gravity}

Since we expect to end up studying a quadratic ETG, it might be helpful to briefly recollect some relevant results for this particular class of higher-order ETG's. The starting point can be considered the general action
\be\label{aquad}
\mathcal{S}_{\text{quadr}} = \frac{1}{2\kappa} \int_{\mathcal{M}} \left[R + \left(\alpha R^{2} + \beta R_{\mu\nu} R^{\mu\nu} + \gamma R_{\theta\iota\kappa\lambda} R^{\theta\iota\kappa\lambda} \right) \right] \, \sqrt{-g} \, \mathrm{d}^{4}x \;,
\ee 
with $\alpha,\beta,\gamma$ three real constants, unspecified for the moment. We already know that we can expect at most $8$ gravitational d.o.f.'s in such a model. We also know that many of such dynamical variables will be ghosts~\cite{Nunez:2004ts,Chiba:2005nz} --- see also our discussion in Sects.~\ref{propag}, \ref{quadgrav}, and~\ref{ghosts}.\footnote{Another way to look at ghosts, at the quantum level at least, is in terms of loss of predictability. Even though an ETG can be made renormalizable by adding quadratic combinations of the Ricci and Riemann tensor~\cite{Stelle:1976gc}, the unitarity of the dynamical evolution gets lost in general. More on this topic in Sect.~\ref{ghosts}.}

When exploring the possible relationships among the three parameters $\alpha,\beta,\gamma$, a few relevant candidate theories emerge; a short list of them goes as follows.
\begin{itemize}
	\item $\alpha=\beta=\gamma=0$. This is just Einstein's GR, with its $2$ propagating degrees of freedom, encoded into a massless spin-$2$ graviton;
	\item $\beta=-4\gamma$. This choice reduces the action to that of $f(R)$-ETG's. Therefore, $3$ gravitational d.o.f.'s are expected, and they can be represented by one massive scalar field and the standard graviton (this in view of the proven equivalence between $f(R)$-gravity and scalar-tensor theories of the Brans--Dicke type);
	\item $\alpha=- (\beta+\gamma)/3$. In this case the extra d.o.f.'s are $5$, all gathered into a massive spin-$2$ field juxtaposing the graviton~\cite{Stelle:1976gc,Stelle:1977ry}. The non-graviton part is Weyl's \emph{poltergeist}, a type of ghost field~\cite{Hindawi:1995an,Chiba:2005nz,Nunez:2004ts};
	\item $\alpha,\beta,\gamma$ unconstrained. This is the most general case, and there will be $8$ degrees of freedom in total; it is still possible to rearrange them so as to give the massless graviton ($2$ d.o.f.'s), the massive spin-$2$ field ($5$) and $1$ scalar field ($1$).
\end{itemize}

Before moving on, one final word of warning: in what follows, we shall assume that no global topological complications occur~\cite{Barth:1983hb,Alonso:1994tr}, i.e. the same set of field equations emerge after using both the Lagrangian in Eq.~\eqref{aquad}, and one where a Gauss--Bonnet term $\mathcal{G}$ has been added to the bulk action. As a consequence~\cite{Barth:1983hb}, we can always trade the Riemann-squared terms in Eq.~\eqref{aquad} for Ricci-squared and scalar-curvature-squared terms. We are also excluding a possible $(\square R)$-term in the action~\eqref{aquad}, as the latter can be recast in the form of a total divergence, hence dynamically irrelevant.

\subsection{The expansion procedure}

Following the steps in Refs.~\cite{Hindawi:1995cu,Chiba:2005nz}, we perform an expansion of Eq.~\eqref{fRiem} up to second order in curvature invariants around a MS solution to study the excitations of the theory around such background. Notice that, in view of the results reported in  Sect.~\ref{propagators}, it is the quadratic terms in the expansion that are the relevant ones to determine the particle content.

It is worth noticing that, regardless of the particular $f (g_{\mu\nu} , R_{\alpha\beta\gamma\delta})$-theory considered, the resulting ``effective'' Lagrangian emerging after the expansion will always be of the quadratic type~\eqref{aquad} --- the major difference will be the specific set of values retrieved for the constants $\alpha,\beta,\gamma$. Hence, all the main features of the ETG's at hand can be studied already at the level of second-order corrections.

For the expansion, we shall consider the MS solutions $g_{\mu\nu}^{(0)}$ such that the two conditions below occur~\cite{Hindawi:1995cu,Chiba:2005nz} 
\be
R_{ab} \equiv R^{(0)}_{\mu\nu} = \frac{R^{(0)}}{4} g_{\mu\nu},
\qquad
R_{\theta\iota\kappa\lambda} \equiv R^{(0)}_{\theta\iota\kappa\lambda}=\frac{R^{(0)}}{12} \left(g_{\theta\kappa} g_{\iota\lambda} - g_{\theta\lambda} g_{\iota\kappa}\right),
\ee
with $R^{(0)}$ the constant value of the scalar curvature for the given background. We can therefore expand the action as
\begin{equation}
\begin{split}
\mathcal{S}_{\text{q}} &=\frac{1}{2 \kappa} \int_{\mathcal{M}} \sqrt{-g}\, \mathrm{d}^{4} x \left\{a_{(0)}+a_{(1)} \left(R - R^{(0)}\right)\right. \\
&\quad\left. +\frac{1}{2} \left[ a_{(2,1)} \left( R - R^{(0)} \right)^{2} + a_{(2,2)} \left( R_{ab} - R^{(0)}_{\mu\nu} \right)^{2} + a_{(2,3)} \left( R_{\theta\iota\kappa\lambda} - R^{(0)}_{\theta\iota\kappa\lambda} \right)^{2} \right]\right\},
\end{split}
\end{equation}
where we have introduced the shorthand notations
\begin{gather}
a_{(0)}=\left.f\right|_{R^{(0)}}, \\
a_{(1)} (R-R^{(0)}) = \left.\dfrac{\mathrm{d} f}{\mathrm{d} R_{\theta\iota\kappa\lambda}}\right|_{R^{(0)}} (R_{\theta\iota\kappa\lambda}-R^{(0)}_{\theta\iota\kappa\lambda}),
\end{gather}
\begin{equation}
\begin{split}
a_{(2,1)} &\left( R - R^{(0)} \right)^{2} + a_{(2,2)} \left( R_{\mu\nu} - R^{(0)}_{\mu\nu} \right)^{2} + a_{(2,3)} \left( R_{\theta\iota\kappa\lambda} - R^{(0)}_{\theta\iota\kappa\lambda} \right)^{2}=\\
& \left.\dfrac{\mathrm{d}^{2} f}{\mathrm{d} R_{\rho\sigma\tau\upsilon} \mathrm{d} R_{\xi\zeta\varsigma\omega}}\right|_{R^{(0)}} \left( R_{\rho\sigma\tau\upsilon} - R^{(0)}_{\rho\sigma\tau\upsilon} \right) \left( R_{\xi\zeta\varsigma\omega} - R^{(0)}_{\xi\zeta\varsigma\omega} \right).
\end{split}
\end{equation}

To further proceed, we make use of the constant curvature condition $a_{(1)}R^{(0)} = 2 a_{(0)}$, which is to say the restriction of the field equations to constant curvature solutions~\cite{Hindawi:1995cu}. By collecting the pieces order by order, we reach the final form of the  action, which reads
\begin{equation}\label{linact}
\begin{split}
\mathcal{S}_{\text{q}} = \frac{\xi}{2 \kappa} \int_{\mathcal{M}} \!\!\sqrt{-g} \, \mathrm{d}^{4} x \left\{-\frac{R^{(0)}}{2} + R + \frac{1}{6 m_{(0)}^{2}} R^{2} -\frac{1}{m_{(2)}^{2}} \left( R_{\mu\nu} R^{\mu\nu} - \frac{1}{3} R^{2} \right) + \zeta \, \mathcal{G}\right\},
\end{split}
\end{equation}
where the four symbols $\xi, \zeta, m_{(0)}$ and $m_{(2)}$ stand for, respectively
\begin{gather}
\xi = a_{(1)} - \left( a_{(2,1)} + \frac{1}{4} a_{(2,2)} + \frac{1}{6} a_{(2,3)} \right) R^{(0)} \;,
\\
m_{(0)}^{2}=\dfrac{\xi}{3 a_{(2,1)} + a_{(2,2)} + a_{(2,3)}},
\\
m_{(2)}^{2} = - \frac{2 \xi}{a_{(2,2)} + 4 a_{(2,3)}}, 
\\
\zeta=\frac{a_{(2,3)}}{2\xi}.
\end{gather}   

Eq.~\eqref{linact} shows that the expanded Lagrangian is nothing but that of a quadratic ETG with an effective cosmological constant term (given by $-R^{(0)}/2$), and a gravitational coupling rescaled by $\xi$. For this theory, as already mentioned multiple times, the number and type of gravitational d.o.f.'s are already known.

What is the lesson learnt from the application of this expansion of the action around MS solutions? Apparently, that the actual dynamical content of \emph{any} fourth-order higher-curvature ETG can be effectively framed and identified by following this method. Not only that: we have also just concluded that the dynamical variables emerged so far are those of a quadratic ETG whose ``weighting coefficients'' are determined by the specific form of the starting action (the dependence of $\xi,\zeta,m_{(0)}$ and $m_{(2)}$ on the elements of the starting action). All in all, quite a satisfying conclusion, and an impressively-performing extraction method.

Unfortunately, the triumphs of the expansion technique are far less effective than what might appear from the discussion above. We shall conclude in a moment that the seemingly far-reaching recipe based on MS solutions and curvature invariants has several structural problems, and does not fare well as it promises to.

\subsection{Problems of the method: general discussion}

The expansion around MS solution is undoubtedly a simple and fast method to study higher-order ETG's around their vacua, yet its imperfections quickly reveal. Serious issues already emerge when examining the case of a relatively minimal $f(R)$-theory defined by $f(R):=R+\chi R^{3}$, with $\chi$ a coupling constant. Indeed, when expanding this action around Minkowski space (which itself is a MS solution), the scalar degree of freedom does not appear, as already discussed in Sect.~\ref{consprop}. At the same time, as soon as we perform the very same expansion around a \emph{non-flat MS solution} (if any), the scalar d.o.f. will eventually crop up.

This last conclusion suggests that the decoupling of the non-metric scalar field is just the occasional effect of the expansion around a single, specific background, with peculiar properties as already discussed in Sect.~\ref{propagators}. The situation gets worse, actually. As we are about to show, there are other cases where the additional d.o.f.'s cannot be made manifest after an expansion around \emph{any} MS solution of the theory. This is a serious wound for the technique, and its biggest limitation.

As a clarifying example, consider a theory defined by the action
\be\label{rxyaction}
\mathcal{S}=\frac{1}{2\kappa}\int_{\mathcal{M}}d^{4}x\sqrt{-g} f\left(R,X,Y\right),
\ee
where $X := R_{\mu\nu} R^{\mu\nu}$ and $Y := R_{\xi\zeta\rho\sigma}R^{\xi\zeta\rho\sigma}$.
The constant-curvature condition characterizing the MS solutions of such model is given by~\cite{Nunez:2004ts}
\be\label{maxsim}
f - \frac{1}{2} R^{(0)} f_{R} - \frac{1}{4} \left(R^{(0)}\right)^2 f_{X} - \frac{1}{6} \left(R^{(0)}\right)^2 f_{Y} = 0,
\ee
where $f_{A_{i}} := \frac{\partial f}{\partial A_{i}}$ with $A_{1,2}=X,Y$, and all the terms are evaluated at $R=R^{(0)}$ (i.e., the value of the scalar curvature fixed for each MS solution). To proceed further, we rewrite Eq.~\eqref{rxyaction} in terms of the trace-free part of the Ricci tensor, denoted by $S_{\mu\nu}$, and the Weyl tensor $C_{\alpha\beta\gamma\delta}$. Since both these tensors vanish for MS spacetimes, the calculations simplify considerably. A few algebraic manipulations lead to
\begin{gather}
R_{\mu\nu} = S_{\mu\nu} + \frac{R^{2}}{4} \,,\label{fir}\\
\begin{split}
R_{\xi\zeta\rho\sigma} R^{\xi\zeta\rho\sigma} &= C_{\alpha\beta\gamma\delta} C^{\alpha\beta\gamma\delta} + 2 R_{\mu\nu} R^{\mu\nu} - \frac{1}{3} R^{2} \\
&= C_{\alpha\beta\gamma\delta} C^{\alpha\beta\gamma\delta} + 2 S^{\mu\nu} S_{\mu\nu} + \frac{1}{6} R^{2},\label{sec}
\end{split}
\end{gather}
where the last equality in Eq.~\eqref{sec} requires Eq.~\eqref{fir} to hold.

Eq.~\eqref{rxyaction} can thus be reformulated as
\be\label{rscaction}
\mathcal{S}=\frac{1}{2\kappa}\int_{\mathcal{M}}\sqrt{-g}\,\mathrm{d}^{4}x f\left(R,W,Z\right),
\ee
where now $W := S^{\mu\nu} S_{\mu\nu}$ and $Z := C_{\alpha\beta\gamma\delta} C^{\alpha\beta\gamma\delta}$. We can then compute the derivatives of $f$ evaluated at $R=R^{(0)}$, namely
\begin{eqnarray}
&\left.\displaystyle\frac{\partial f}{\partial S_{\mu\nu}}\right|_{R^{(0)}}=\left.\left(\displaystyle\frac{\partial f}{\partial W} \displaystyle\frac{\partial W}{\partial S_{\mu\nu}}\right)\right|_{R^{(0)}}=0 &
\\
&\left.\displaystyle\frac{\partial f}{\partial C_{\xi\zeta\rho\sigma}}\right|_{R^{(0)}}=\left.\left(\displaystyle\frac{\partial f}{\partial Z}\displaystyle\frac{\partial Z}{\partial C_{\xi\zeta\rho\sigma}}\right)\right|_{R^{(0)}}=0, &
\end{eqnarray}
where we used the vanishing of $C_{\alpha\beta\gamma\delta}$ and $S_{\mu\nu}$ on MS solutions. It can be shown, in the same way, that all the mixed derivatives vanish as well.

The non-vanishing derivatives are instead given by
\begin{eqnarray}
&\displaystyle\frac{1}{2}\left.\displaystyle\frac{\partial^{2} f}{\partial S_{\iota\kappa} S_{\mu\nu}}\right|_{R^{(0)}}S_{\iota\kappa}S_{\mu\nu}=\displaystyle\frac{1}{2} \left.\displaystyle\frac{\partial f}{\partial W}\right|_{R^{(0)}} W & \\
&\displaystyle\frac{1}{2}\left.\displaystyle\frac{\partial^{2} f}{\partial C_{\alpha\beta\gamma\delta}C_{\xi\zeta\rho\sigma}}\right|_{R^{(0)}}C_{\alpha\beta\gamma\delta}C_{\xi\zeta\rho\sigma}=\left.\displaystyle\frac{\partial f}{\partial Z}\right|_{R^{(0)}} Z. &
\end{eqnarray}
If now we keep only the terms which are constant, linear and quadratic in the curvature, we arrive at
\begin{align}\label{linChiba}
\mathcal{S}_{q} =\frac{1}{2\kappa} \int_{\mathcal{M}} &\sqrt{-g}\,\mathrm{d}^{4}x \left\{f(R^{(0)})+\left.\frac{\partial f}{\partial R}\right|_{R^{(0)}}(R-R^{(0)})+\frac{1}{2}\left.\frac{\partial^{2} f}{\partial R^{2}}\right|_{R^{(0)}}(R-R^{(0)})^{2}\right.  \nonumber \\
& \left.+\left.\frac{\partial f}{\partial W}\right|_{R^{(0)}} W +\left.\frac{\partial f}{\partial Z}\right|_{R^{(0)}}Z\right\}  \nonumber \\
=\frac{1}{2\kappa}\int_{\mathcal{M}}&\sqrt{-g}\,\mathrm{d}^{4}x \left\{\left(f(R^{(0)})-\left.\frac{\partial f}{\partial R}\right|_{R^{(0)}}R^{(0)}-\left.\frac{\partial^{2} f}{\partial R^{2}}\right|_{R^{(0)}}\left(R^{(0)}\right)^{2}\right)\right. \nonumber \\
&\left.+\left(\left.\frac{\partial f}{\partial R}\right|_{R^{(0)}}-\left.\frac{\partial^{2} f}{\partial R^{2}}\right|_{R^{(0)}}\right)R +\left(\frac{1}{2}\left.\frac{\partial^{2} f}{\partial R^{2}}\right|_{R^{(0)}}+\frac{1}{12}\left.\frac{\partial f}{\partial W}\right|_{R^{(0)}}\right)R^{2}\right. \nonumber \\
&\left. +\left(\left.\frac{\partial f}{\partial Z}\right|_{R^{(0)}}+\frac{1}{2}\left.\frac{\partial f}{\partial W}\right|_{R^{(0)}}\right)Z- \left.\frac{\partial f}{\partial Z}\right|_{R^{(0)}}\mathcal{G}\right\}, 
\end{align}
where in the second equality we have massaged the expression to single out the Gauss--Bonnet combination, using the relation~\cite{Chiba:2005nz}
\begin{equation}\mathcal{G}=C_{\alpha\beta\gamma\delta} C^{\alpha\beta\gamma\delta}-2S^{\mu\nu}S_{\mu\nu}+\frac{1}{6}R^{2}\end{equation}

We can now discuss the d.o.f.'s of the resulting theory by looking at the quadratic action in eq.~\eqref{linChiba} and using the knowledge about the d.o.f.'s in any quadratic ETG. To better grasp the connection, note that Eq.~\eqref{aquad} can be rewritten, using the Gauss--Bonnet identity, in terms of $C_{\alpha\beta\gamma\delta} C^{\alpha\beta\gamma\delta}$ and $R^2$ as
\be\label{aquad2}
\mathcal{S}_{q}=\frac{1}{2\kappa}\int_{\mathcal{M}}\sqrt{-g}\,\mathrm{d}^{4}x\left[R+\frac{1}{6m_{0}^{2}} R^{2}-\frac{1}{2m_{2}^{2}} C_{\alpha\beta\gamma\delta} C^{\alpha\beta\gamma\delta}\right].
\ee

The presence of the scalar and spin-$2$ massive fields is related to the presence of the $R^{2}$- and the Weyl-squared terms in the quadratic action, respectively. It is possible to prevent these d.o.f.'s from appearing by simply picking the right values for the quantities multiplying $R^{2}$ and $C_{\alpha\beta\gamma\delta} C^{\alpha\beta\gamma\delta}$ in Eq.~\eqref{linChiba}. The general requirements for a theory not to present both the non-metric scalar and poltergeist d.o.f.'s read
\be\label{cond}
\left\{
  \begin{array}{l l}
    2\displaystyle\frac{\partial f}{\partial Z}=-\displaystyle\frac{\partial f}{\partial W}, & \quad \text{No extra spin-2 field}\\[0.8em]
    6\displaystyle\frac{\partial^{2} f}{\partial R^{2}}=-\displaystyle\frac{\partial f}{\partial W}, & \quad \text{No extra scalar field,}
  \end{array}\right.
\ee
where all the derivatives are evaluated at $R^{(0)}$. If both these conditions are satisfied, neither the spin-2 field nor the scalar field can show up in the quadratic action.

As a result, we conclude that the expansion method does not allow to distinguish, in general, between theories that do not have non-metric d.o.f.'s at the full non-linear level, from theories that simply do not make such d.o.f.'s emerge around MS spacetimes. The same arguments apply for the action~\eqref{linact}, which encompasses even more general theories. The only difference, in this latter case, is that it is not as easy to find the general form of Eqs.~\eqref{cond} in more involved contexts.

\subsection{Problems of the method: examples}

To further reinforce our criticism against the expansion method, we now show two explicit examples where the MS-expansion fails at providing the required information about the dynamical content of ETG's.

We shall build explicitly a class of $f(R,\mathcal{G})$-theories which, albeit endowed with an extra scalar d.o.f., do satisfy  the second condition in Eq.~\eqref{cond}. Noticeably, for this class of theories the first condition in eq.~\eqref{cond} is always satisfied, in accordance with the fact that in $f(R,\mathcal{G})$-ETG's, no extra spin-$2$ field is expected at the full non-linear level --- see Chapter 12 in~\cite{DeFelice:2010aj} and references therein. The action we are interested in has the form
\be
\mathcal{S}=\frac{1}{2\kappa}\int_{\mathcal{M}}R-\alpha\left[\frac{1}{108}R^{4}-\frac{1}{2}\mathcal{G}^{2}\right],
\ee
where $\alpha$ is an otherwise unspecified constant of mass dimension $-2$. Since such ETG gives field equations at fourth order, we expect one extra scalar d.o.f., as the model under scrutiny indeed reveals. In terms of the previously-introduced objects $R, W, Z$, the Lagrangian density is given by
\be
f(R,W,Z)=R-\frac{\alpha}{108}R^{4}+\frac{\alpha}{2}\left(Z-2W+\frac{1}{6}R^{2}\right)^{2}.
\ee 
and we have 
\be
\left.\frac{\partial^{2} f}{\partial R^{2}}\right|_{R^{(0)}}=\frac{\alpha}{18}\left(R^{(0)}\right)^{2} \qquad \left.\frac{\partial f}{\partial W}\right|_{R^{(0)}}=-\frac{\alpha}{3}\left(R^{(0)}\right)^{2} \qquad \left.\frac{\partial f}{\partial Z}\right|_{R^{(0)}}=\frac{\alpha}{6}\left(R^{(0)}\right)^{2}. 
\ee

From these expressions, it is easy to realize that both the conditions in Eq.~\eqref{cond} are satisfied. The expanded action contains only $R$ and $\mathcal{G}$ at order $2$ in the curvatures (except for a possible non-zero cosmological constant term), and hence no extra d.o.f.'s are present. Also, using Eq.~\eqref{maxsim} and the two relations $X=\frac{R_{0}^{2}}{4}$, $Y=\frac{R_{0}^{2}}{6}$ (which hold for MS solutions), it is possible to show that indeed the theory admits two MS solutions, i.e. Minkowski and de Sitter (for the latter, it is $R^{(0)}=\sqrt[3]{108/\alpha}$).

Therefore, this case is qualitatively different from the example of a cubic $f(R)$-theory, since the extra d.o.f. (expected on general grounds) does not show up around \emph{any} MS solution. As a result, in such ETG the scalar d.o.f. is completely hidden beneath the level of the expansion method, and the latter cannot provide any useful bit of information as per the existence of the scalar, which is a proper dynamical ingredient of the model, even though it does not manifest as a perturbation on MS spacetimes.

As a second example, we exhibit a suitably-formulated ETG such that it is the spin-$2$ extra d.o.f. that is always absent on MS spacetimes.  
The Lagrangian density is given by 
\be
f(R,X,Y)=R+\alpha\left\{\exp\left(-\frac{2}{3} \frac{1}{\Upsilon} X\right)-\frac{1}{6}\exp\left(-\frac{1}{\Upsilon} Y\right)\right\},
\ee
where $\alpha$ and $\Upsilon$ are two constants with mass dimension $-2$ and $X,Y$ are defined right below Eq.~\eqref{rxyaction}. By general arguments, both a spin-$2$ and a scalar non-metric d.o.f.'s should be expected. Quite the opposite, instead, the expanded action does not provide any extra spin-$2$ field, since the first of the conditions~\eqref{cond} is identically satisfied. Once again, one of the potentially expected non-metric fields seems to be ``invisible'', with the method presented here insensitive to its eventual existence --- we say that it \emph{seems} because the lack of a full (and method-independent) analysis of this particular ETG does not allow to conclude whether or not the additional d.o.f.'s not emerging from the expansion can be finally made resurrect.

\subsection{Considerations on the method}

Once again, we face an extraction technique which leads to a few improvements, but soon after appears to be inconclusive at best, and proves to be unable to fully determine the dynamical content of a reasonable sub-set of the higher-order ETG's.

More specifically, we have shown that the expansion around maximally symmetric spacetimes of $f(g_{\mu\nu}, {R^\alpha}_{\beta\gamma\delta})$-ETGs, while incorporating elements from both the linearization/propagator procedure, and the auxiliary-fields recipe, does not work as hoped, and fails at correctly profiling the ETG's whenever an extra d.o.f. falls below the scanning threshold of the MS solutions themselves.
	
At this point, it may be fair to draw a broader conclusion about the results achieved so far. The main message here is that one ought to avoid relying on the various methods discussed up to now to make any bold statement on the actual dynamical content of higher-order ETG's, unless some other independent source of information is available. Luckily enough, there are better ways to deal with gravity theories (and field theories in general), even though the price one pays for a better understanding of the nature of the d.o.f.'s is a significant growth in the computational complexity.%
\section{Hamiltonian formalism \label{hamiltonian}}

After presenting and critically discussing a certain number of methods frequently used to extract the gravitational d.o.f.'s in higher-order ETG's it is about time to move to the most effective and powerful tool to dissect a field theory and unveil its actual dynamics: the \emph{Hamiltonian formalism}. This protocol pins down the relevant features of a given model for physical phenomena, and works out its evolution equations by incorporating in the treatment any supplementary bit of information one can gain by manipulating the action, the equations of motion, and the symmetries.

The Hamiltonian formalism is in fact the state-of-the-art method to investigate ETG's (and any other physical theory, actually). Even though it is not perfect, and its problems are sometimes very well hidden, it is overwhelmingly more powerful than anything seen so far, and any attempt at describing the ways to tackle a theory of gravity would not be complete without a proper mention of such protocol. 

In what follows, we will not dive deep into technical details; rather, we shall try instead to review the achievements, and then the subtleties and main drawbacks of the recipe, highlighting the elements which make the Hamiltonian formalism at once the most appealing tool for the analysis of ETG's, but also one of the most difficult to master.

For a more thorough exploration of this method, we first and foremost cite the seminal works by Dirac himself~\cite{Dirac:1950pj,dirac2013lectures}; then, the authoritative book~\cite{Henneaux:1992ig} makes for a complete treatise on the topic. An enlightening yet somewhat tangential perspective --- but one with a strong connection with GR and the innermost structure of ETG's --- is offered by the Shape Dynamics tutorial~\cite{Mercati:2014ama}.

\subsection{Constrained Hamiltonian systems and the Dirac formula}

Counting the actual degrees of freedom in a dynamical model is in general a nontrivial task; the situation gets rapidly worse whenever the theory at hand enjoys some sort of local symmetry, as it occurs in most descriptions of fundamental physical phenomena. GR, to give an example, is reparametrization-invariant for spacetime diffeomorphisms, and the same can be said for most ETG's implemented via a covariant formulation.

That local symmetries are a feature of a physical theory can be inferred by the presence, in the solutions of the equations of motion, of arbitrary functions of time. Such feature is tantamount to saying that the description of the dynamical evolution has been made in terms of redundant variables, and that we are not pinning down the actual d.o.f.'s. A third way to frame the problem is to state that the model exhibits a form of gauge invariance~\cite{baez-relknots}: the physics is not affected by a specific gauge choice, whereas the description of interactions and evolution does change with the gauge.

To recover the ``true'' number of actual d.o.f.'s, it is necessary to implement an apt set of \emph{constraints} in the formal machinery of the theory, which supplement the standard Hamiltonian picture. The equations for the constraints combine the canonical coordinates together and compensate the redundant initial choice for the dynamical variables, reducing the number of independent functions. Therefore, gauge theories must be treated as constrained systems~\cite{Henneaux:1992ig}.

The Hamiltonian formalism for constrained systems was developed in the early Fifties by Dirac~\cite{Dirac:1950pj,Dirac:1958sq} (see also Refs.~\cite{dirac2013lectures, Anderson:1951ta,Bergmann:1954tc}) and further developed in the following years (additional references in~\cite{Henneaux:1992ig}). The protocol is widely recognized as the most thorough and foolproof way to deal with such systems. In what follows, we provide a very basic introduction to the foundations of such formalism, the goal being a justification of the famous Dirac formula counting the physical d.o.f.'s of gauge theories.

As the topic somewhat deviates from the main context and tools one is familiar with in the study of ETG's, we prefer to elaborate for a few paragraphs on the steps required to get to the final results.

\subsubsection{From Lagrange to Hamilton}

An elegant and compact way to display the information about a physical theory is to build up the associated action functional --- notations and conventions adapted from~\cite{Henneaux:1992ig} --- which typically reads
\be
\mathcal{S} [q^I (t)]=\int_{t_{0}}^{t_{1}} \mathrm{d}t\, \mathcal{L}(q^I,\dot{q}^I),
\ee  
where the $q^I$'s for $I=1,2,\dots, N$ are the Lagrangian coordinates, and the Lagrangian function $\mathcal{L}(q^I,\dot{q}^I)$ is, at the classical level, a difference of kinetic and potential terms. A curve $\bar{q}^I (t)$ on the coordinate space is a \emph{classical trajectory} if the action is stationary when evaluated on such a curve, that is if $\delta \mathcal{S} [\bar{q}^I (t)] = 0$ for all variations $\delta q^{I}$ vanishing at the endpoints $(t_{0}, t_{1})$.

The variational problem of the search for the extrema of $\mathcal{S}$ implies that the following Euler--Lagrange equations of motion hold for the $q^I$'s, i.e.
\be\label{eulerlagr}
\ddot{q}^I \frac{\partial^{2}\mathcal{L}}{\partial \dot{q}^I\partial \dot{q}^J}=\frac{\partial \mathcal{L}}{\partial q^J}-\dot{q}^I\frac{\partial^{2} \mathcal{L}}{\partial q^I\partial \dot{q}^J} \,.
\ee 
The previous system can be brought into its \emph{normal form} --- that is, the $\ddot{q}^I$ can be uniquely determined by the initial conditions $q^I (t_0)$ on positions and $\dot{q}^I (t_0)$ on velocities --- if and only if it occurs that
\be\label{regular}
\det\left(\frac{\partial^{2} \mathcal{L}}{\partial \dot{q}^I\partial \dot{q}^J}\right)\neq 0.
\ee
The fulfillment of such conditions is the defining feature of the so-called \emph{regular Lagrangians}; if instead Eq.~\eqref{regular} is not satisfied, the corresponding Lagrangian $\mathcal{L}$ is said to be \emph{singular}.

For regular Lagrangians, the second-order system~\eqref{eulerlagr} can be easily remapped into a first-order form according to a procedure due to Hamilton; upon defining the conjugate momenta $p_J$ as
\be\label{momenta}
p_J := \frac{\partial \mathcal{L}}{\partial \dot{q}^J},
\ee
and introducing the \emph{canonical Hamiltonian} function $\mathcal{H} (q^I , p_J)$ --- in fact, the Legendre transform of $\mathcal{L}$ with respect to the $\dot{q}^I$'s --- given by
\begin{equation}\label{hamildef}
\mathcal{H} (q^I , p_J) := \dot{q}^I p_I - \mathcal{L}
\end{equation}
then the Euler--Lagrange equations of motion equal Hamilton's equations
\begin{equation}\label{hamiltoneom}
\dot{q}^I = \frac{\partial \mathcal{H}}{\partial p_I} \qquad,\qquad \dot{p}_J = - \frac{\partial \mathcal{H}}{\partial q^J} \,,
\end{equation}
which are defined on the $2N$-dimensional product manifold of the coordinate and momentum spaces, i.e. the phase space $\varGamma$, where a symplectic structure is induced and preserved by the evolution. A classical trajectory is then a new curve on $\varGamma$ solving equations~\eqref{hamiltoneom} for given initial positions and momenta.

Hamilton's equations~\eqref{hamiltoneom} allow for the identification of the dynamics of any function defined on $\varGamma$; indeed, if $f = f (q,p) = f (t)$, then its evolution can be written as
\begin{equation}\label{eomgenericf}
\frac{\partial f}{\partial t} = \frac{\partial f}{\partial q^I} \frac{\partial q^I}{\partial t} + \frac{\partial f}{\partial p_J} \frac{\partial p_J}{\partial t} = \frac{\partial f}{\partial q^I} \frac{\partial \mathcal{H}}{\partial p_I} - \frac{\partial f}{\partial p_J} \frac{\partial \mathcal{H}}{\partial q^J} =: \left\{ f , \mathcal{H} \right\}
\end{equation}
where the third term contains the general definition of the \emph{Poisson brackets} $\left\{ f , g \right\}$ between two functions on the phase-space.

The present construction, initially conceived for discrete-particles systems, can be seamlessly rewritten for continuous fields defined over $m$-dimensional manifolds, leading to the Lagrangian and Hamiltonian formulation of field dynamics for actions of the type $\mathcal{S} \left[ \Phi^I (x^\alpha) , \partial_\beta \Phi^I (x^\alpha) \right]$.

\subsubsection{Singular Lagrangians and constrained systems}

Whenever condition~\eqref{regular} is not satisfied and the Lagrangian function $\mathcal{L}$ is singular, a specific complication occurs: the (generalized) velocities $\dot{q}^I$'s cannot be defined anymore uniquely in terms of the coordinates $q^I$'s and the momenta $p_J$'s; stated otherwise, not all the $p_J$'s are independent, but some of them can be expressed as combinations of the coordinates, via the relations 
\be\label{primconstr}
\phi_{m}(q,p)\approx 0, \ m=1,\dots, M
\ee
which are called \emph{primary constraints} --- ``primary'' because their existence is prior to the solutions of the equations of motion --- and depend on the singular nature of the corresponding Lagrangian. The number $M$ accounting for the primary constraints is given by
$$M=N-\rank \left(\frac{\partial^{2}L}{\partial \dot{q}^{I}\partial \dot{q}^{J}}\right) \;.$$ 

The symbol ``$\approx$'' on the left of Eq.~\eqref{primconstr} must be read as \emph{weakly equal}, and means that the vanishing of the functions $\phi_m$'s is restricted to the actual solutions of the equations of motion, and not on the whole phase space. In other words, the $\phi_m$'s define a submanifold of $\varGamma$, and their Poisson brackets with the canonical variables is in general different from zero. As a consequence, any two functions $\mu (q,p) , \nu (q,p)$ on phase space which coincide on the constraints submanifold are said to be weakly equal, and we can write $\mu \approx \nu$.

The presence of primary constraint in a given theory also affects the construction~\eqref{hamildef} of the canonical Hamiltonian, which is not defined uniquely anymore on $\varGamma$.\footnote{It gets worse than that: general-covariant theories such as the ETG's we are dealing with result in general in a Hamiltonian function $\mathcal{H}$ which is itself a constraint, i.e. $\mathcal{H} \approx 0$. This lies at the heart of the so-called ``problem of time'' in quantum gravity~\cite{Anderson:2012vk}.} The problem can be fixed by introducing the so-called \emph{Dirac Hamiltonian}, which takes the constraints into account without modifying the global canonical formalism. The new Dirac Hamiltonian is a linear combination of $\mathcal{H}$ and all the primary constraints, reading
\be
\mathcal{H}_{D} := \underbrace{\dot{q}^{n}p_{n}-L}_{\mathcal{H}_{\text{can}}}+\sum_{i=1}^{M}\lambda^{m}(q,p) \phi_{m} \approx \mathcal{H}_{\text{can}},
\ee
where the $\lambda^{m}(q,p)$'s can be seen as Lagrange multipliers enforcing the primary constraints upon variations of the action. The $\mathcal{H}_D$ can be used, in place of Eq.~\eqref{eomgenericf}, to establish once again the dynamics of any function $f (q,p)$ on $\varGamma$, resulting in $\partial_t f = \left\{ f , \mathcal{H}_D \right\}$. At the same time, by consistency, primary constraints must be preserved by time evolution, i.e. for all $m$ it must be
\be\label{pres}
\left\{\phi_{m},\mathcal{H}_{D}\right\}\approx 0.
\ee
The system of equations above leaves us with only three possibilities: i) the equations are trivially satisfied; ii) the equations can determine the values of the Lagrange multipliers $\lambda^{m}(q,p)$'s; iii) the equations are themselves independent from the $\lambda^{m}(q,p)$'s, and as such they yield a new set of further constraints. In this latter case, we can say we have found the  \textit{secondary constraints} of the theory.

The protocol sketched above can result in a hierarchy of constraints: every time a new level emerges, one has to check which of the three possibilities just stated occurs at that level, until all consistency relations are fulfilled.\footnote{Already when secondary constraints appear, one also has to take care of all the relations emerging from the deployment of the Lagrange multipliers. By properly massaging the sets of equations, one can define finally a \emph{total Hamiltonian}~\cite{Henneaux:1992ig}, which will contain a certain number of arbitrary functions of time. Such functions will propagate into the solutions of the equations of motion, confirming the underlying presence of a gauge symmetry (or, which is the same, of a certain type of constraints --- see below).}

Once we have all the constraints at our disposal, the next step is to classify them in \textit{first-class} and \textit{second-class} constraints. This is indeed the main step we need to take to finally build up a formula to count the d.o.f.'s of a given theory.

\subsubsection{Dirac's formula for the d.o.f.'s}

A function $f(q,p)$ on phase space is called a \emph{first-class function} if its Poisson bracket with every constraint vanishes weakly, i.e. if $\{f,\phi_m\} \approx 0$ for all $m$. Whenever such condition is not satisfied, $f$ is called a \textit{second-class function}. The first-class property is such that, if $f$ and $g$ are first-class functions, so is the combination $\{f,g\}$. By solving the preservation conditions~\eqref{pres} involving Lagrange multipliers --- i.e., falling under case ii) outlined above --- one is able to find the number of independent first-class primary constraints.

Primary first-class constraints are widely assumed to be the \emph{generators of gauge transformations}, and the latter do not alter the physical state of a system, but merely reflect the presence of a redundancy in its description. Still, this conclusion about the role of primary-class constraints within gauge transformations might deserve more carefulness, as there seems to be arguments against its full validity~\cite{Pons:2004pp,Pitts:2013uxa,Pitts:2014nba}.\footnote{More specifically, issues seem to arise in both electromagnetism and Yang--Mills theory, and even in an aptly truncated theory mimicking the features of GR. 
For all the technical details, the Reader is invited to check the cited References.} That all first-class constraints are in the end generators for gauge transformations is the content of \emph{Dirac's conjecture}~\cite{Henneaux:1992ig}.

Since the first-class constraints generate gauge transformations, we are free to define the so-called \emph{extended Hamiltonian}, which takes into account all first-class constraints, is weakly equal to Dirac's Hamiltonian, and gives the same canonical equations as the original action. As for second class constraints, they do not generate any physically important transformation.\footnote{The solution of second-class constraints is achieved by means of the Dirac brackets, used in the equations in lieu of Poisson brackets~\cite{Henneaux:1992ig} --- see two paragraphs below the footnote mark.}

Once we have all first- and second-class constraints, we can deal with the former group by imposing \emph{gauge-fixing conditions} to get rid of the ambiguities in the representation of the actual dynamics of a system, because we can be sure that the physics will not change, regardless of our choice for the gauges. The task is achieved as soon as we set as many gauge fixings as the number of independent first-class constraints --- call this figure $\mathcal{N}_{\mathrm{I}}$. Such number must be subtracted from the number $N$ of initial conditions necessary to specify a solution of the field equations.

Having settled the first-class constraints, we are left with the second-class ones, and those can be accounted for by introducing the so-called \emph{Dirac brackets}~\cite{Henneaux:1992ig}, symbol $\{f,g\}^\ast$; these are modified Poisson brackets (weakly equal to the latter) incorporating the independent primary first-class constraints in the form
\begin{equation}
\{f,g\}^\ast := \{f,g\} - \{f,\phi_i\} C^{ij} \{\phi_j,g\} \,,
\end{equation}
with $C^{ij}$ the inverse matrix of the combination $C_{ij} := \{\phi_i,\phi_j\}$, and the indices $i,j$ spanning the range of independent first-class constraints.

The counting of the actual d.o.f.'s in a theory is completed by finding the number of canonical variables which are solutions of the equations of motion, and satisfy the constraint equations as well. The calculation is condensed in a celebrated formula named after Dirac:
\be\label{dirac}
2 \mathcal{N} = 2 N - \mathcal{N}_{\mathrm{II}} - 2 \mathcal{N}_{\mathrm{I}} \,,
\ee
where $2 N$ is the total number of canonical variables (the $q$'s and the $p$'s, i.e. twice the number of Lagrangian coordinates), and $\mathcal{N}_{II}$ is the number of independent second class constraints.\footnote{It is possible to prove~\cite{Henneaux:1992ig} that $\mathcal{N}_{\mathrm{II}}$ is in general an even number, so Dirac's formula can be also stated as $\mathcal{N} = N - \mathcal{N}_{\mathrm{I}} - \mathcal{N}_{\mathrm{II}}/2$.} The duplication of the first-class constraints ($2 \mathcal{N}_{\mathrm{I}}$) can be seen to emerge from the relation $2 \mathcal{N}_{\mathrm{I}} = \mathcal{N}_{\mathrm{I}} + \mathcal{N}_{\text{gauge}}$, (the so-called ``Dirac conjecture'').

\subsection{Higher-derivative theories: Ostrogradski's algorithm}

The conclusions emerged so far pertain to Lagrangians such that they host at most first derivatives of the Lagrangian coordinates. To properly treat higher-order ETG's we have to adapt the Hamiltonian analysis to the more complex case of theories admitting higher derivatives of the $q$'s in the action, and hence requiring larger sets of initial conditions in the resulting Euler--Lagrange equations. This improvement has been achieved by Ostrogradski some $150$ years ago~\cite{ostrogradski1850member}. The key point is being able to handle the higher time derivatives rather than the spatial ones --- the latter can be regarded as dependent variables once a specific non-covariant description of the dynamics is adopted, see below Sect.~\ref{admdescr}.

In higher-derivative Lagrangians, be they associated with ETG's, or with descriptions of other interesting physical systems~\cite{0305-4470-31-33-006,0305-4470-22-10-021,2013arXiv1305.6744A}, the maximal time derivatives drive the dynamical evolution; at the same time, their presence usually generates chronically unstable models, as the Hamiltonian becomes a linear combination of the canonical momenta, and therefore ends up being often unbounded. However, this last result, a well-known theorem due once again to Ostrogradski~\cite{ostrogradski1850member}, is in fact associated to \emph{regular} higher-derivative Lagrangians, so it does not apply to the singular cases (as e.g. the ETG's we are examining in this work).\footnote{There is more to say on the issue of instability, and we shall briefly comment on it in Sect.~\ref{ghosts}.}

Ostrogradski's reduction method for regular higher-derivative Lagrangians is based on the introduction of new canonical variables such that one falls back into the ordinary case where the highest time derivative appearing in the action is of order one. We start with the original Lagrangian\footnote{We assume for simplicity that for every canonical variable $q^{I}$ the highest derivative is of the same order $k$. Moreover, the regularity condition on the Lagrangian reads: $\rank\frac{\partial^{2}L}{\partial q^{k}_{I}\partial q^{k}_{J}}=N$.}
\be
\mathcal{L} \left( q^{I},\partial_{t} q^{I},\dots,\partial_{t}^{(k)} q^{I} \right),\  I=1,\dots N\ \ k>1\,.
\ee
Then we introduce a new set of Lagrangian coordinates $Q^{J_\ell}$, with $\ell = 1,\dots,k$ via the definition
\be
Q^{J_1}:= q^{I} \,, \: Q^{J_2}:=\partial_{t} q^{I} \,, \dots, \: Q^{J_k}:=\partial_{t}^{(k-1)} q^{I} \;.
\ee
In this way we arrive at the new Lagrangian $\mathcal{L}'$, free of higher derivatives, related to $\mathcal{L}$ by
\begin{equation}
\mathcal{L} \left( q^{I},\partial_{t} q^{I},\dots,\partial_{t}^{(k)} q^{I} \right) = \mathcal{L}' \left( Q^{J_1},Q^{J_2},\dots,Q^{J_k},\dot{Q}^{J_k} \right) \,.
\end{equation}

We can now use the two Lagrangians above interchangeably, so we can define the canonical momenta $P_{J}$ corresponding to the $Q^{J_\ell}$'s as
\be
P_{I}^{(k)}:=\frac{\partial L}{\partial\left(\partial_{t}^{(k)} q^{I}\right)} \qquad,\qquad P^{(h)}_{I}=\frac{\partial L}{\partial\left(\partial_{t}^{(h)} q^{I}\right)}-\partial_{t}P^{(h+1)}_{I},
\ee
where $h=k-1,\dots ,1$. 

The recipe to deal with higher-derivative regular Lagrangians has been extended to the singular case~ \cite{Buchbinder:1986wka,Buchbinder:1987vp, Gitman:1986yb, Pons:1988tj} (see also~\cite{buchbinder1992effective} and references therein). When dealing with singular Lagrangians, one introduces again new variables to encode the higher time-derivatives, but in the process there is a degree of arbitrariness in the choice of the additional variables in view of the possible reformulation of the constraints. As a result, one is left with families of different Hamiltonians, each one emerging from a separate choice of the canonical coordinates. Still, all such families are connected by canonical transformations, which guarantees dynamical equivalence, at least at classical level --- quantization procedures can break the connection between the various formulations~\cite{Buchbinder:1987vp}.

The introduction of the new variables allows to recognize primary (second-class) constraints arising directly from the construction, and this is a desirable feature of the protocol. After these constraints have been set aside, it is possible to proceed to the identification of other primary constraints, and from this step onwards the usual algorithm for gauge theories can be applied.


\subsection{D.o.f.'s in higher-derivative ETG's}

Higher-derivative ETG's enjoy the same diffeomorphisms-invariance germane to GR, and hence are themselves constrained theories. It makes sense, then, to study their dynamics using the Hamilton--Dirac formalism. Their very nature, however, immediately posits a challenge: being higher-derivative theories, they must be first reduced to first-order ones by applying Ostrogradski's method. Such reduction can be achieved in many ways, and smart choices can be made to render the calculations as manageable as possible.

To begin with, we focus on quadratic ETG's, for which a whole lot of results have already been listed in Sect.~\ref{linearization} --- see also Sect.~\ref{Auxiliary fields} and~\cite{Hindawi:1995an} for a different, possibly simpler approach.

On quadratic corrections to GR in a Hamiltonian framework, we shall mainly follow Refs~\cite{buchbinder1992effective,Kluson:2013hza}; for results concerning more general ETG's, one can also peruse the recent contribusions~\cite{Deruelle:2009zk,Sendouda:2011hq}, where the first steps towards a general formulation of the protocol can be found.

\subsubsection{Choice of the ADM variables \label{admdescr}}

The Hamiltonian analysis of an ETG is typically conducted in a special system of coordinates, the ADM variables~\cite{Arnowitt:1962hi,Dirac:1958sc}; even though not mandatory --- generally covariant formulations are possible, see Ref~\cite{LeeWaldcf} --- the introduction of the ADM coordinatization of the spacetime manifold greatly helps the emergence of both the d.o.f.'s and the constraints.

The first step is to assume that the spacetime $(\mathcal{M} , g_{\mu\nu})$ is globally hyperbolic, and hence it can be represented as a stack of three-dimensional, spacelike, non-intersecting Cauchy hypersurfaces $\Sigma$'s, which in turn define a field of surface-orthogonal timelike vectors $n^\alpha$. The different leaves of the foliation are identified via the values assumed by a scalar field $t(x^{\mu})$ defined over $\mathcal{M}$ ($t$ is in fact a continuous, monotonic index labelling the $\Sigma$'s). One is free to further assign three ``spatial'' coordinates $y^i$, $i=1,2,3$ on each and every $\Sigma_t$ in such a way that a displacement along $n^\mu$ does not alter the value of the $y^i$'s.

The spatial character of the $\Sigma_t$'s means that the gradient $\nabla_{\! \mu} t$ generates a timelike vector $\xi^\mu$ via the relation $\xi^\mu \nabla_{\! \mu} t = 1$; now, $\xi^\mu$ can be decomposed as the vector sum of a timelike part (proportional to $n^\mu$), plus a spacelike part orthogonal to it.\footnote{The vector $\xi^\mu$ can be considered the velocity of a congruence of (Eulerian) observers crossing the $\Sigma_t$-leaves and using the $t$-coordinates as an evolution parameter. Since in general one does not assume that the world-lines of the $\xi$-observers pierce the $\Sigma_t$'s orthogonally, a purely spatial contribution $N^\mu$ is needed to match the two displacements given by $n^\mu$ and $\xi^\mu$ along the stack of leaves.} The resulting relation is $\xi^\mu = N n^\mu + N^\mu$, with $N$ ensuring the unit norm for $n^\mu$.

The coordinate system given by the $4$-tuple $y^{\mu} \equiv (t,y^{i})$ just built is adapted to the ADM-representation of $(\mathcal{M} , g_{\mu\nu})$, and any relevant geometric $4$-D quantity can be re-expressed in terms of contributions tangent and orthogonal to the leaves by application of the projectors $\Theta^\mu_\nu := n^\mu n_\nu$ (perpendicular to the $\Sigma_t$'s) and $\Omega^\mu_\nu := \delta^\mu_\nu - n^\mu n_\nu$ (parallel to the stack); this means that the metric $g_{\mu\nu}$ itself can be projected along the leaves of the foliation, and this induces a purely spatial, $3$-D metric $h_{\mu\nu} := \Omega^\alpha_\mu \Omega^\beta_\nu g_{\alpha\beta}$ on each $\Sigma_t$.

It is customary to use different indices (Latin ones, in this case) for the purely spatial quantities, hence the induced metric will be usually written as $h_{ij}$, and also $N^\mu \equiv N^i$ because $N^\mu n_\mu = 0$ and therefore $N^\mu \equiv (0, N^i)$ in the ADM coordinates. The induced metric $h_{ij}$, the lapse function $N$, and the shift vector $N^i$ --- the last two names are self-explanatory --- are the new variables encoding the gravitational d.o.f.'s, as one can check that it is
$$ 
\mathrm{d} s^{2} = g_{\mu\nu} \mathrm{d} x^{\mu} \mathrm{d} x^{\nu} = - N^2 \mathrm{d} t^2 + h_{ij} (N^{i} \mathrm{d} t + \mathrm{d} y^{i}) (N^{j} \mathrm{d} t + \mathrm{d} y^{j}),
$$
with the two determinants $h$ and $g$ related by the relation $\sqrt{-g}=N\sqrt{h}$.

The spatial metric $h_{ij}$ is the first fundamental form on the hypersurfaces $\Sigma_t$'s, and characterizes the intrinsic geometry of the leaves. From $h_{ij}$ and its associated Levi-Civita connection, it is possible to construct the $3$-D curvature tensor, and define the $3$-metric-compatible covariant derivative ($D_i$) on the manifolds $(\Sigma_{t},h_{ij}).$ As for the extrinsic geometry of the $\Sigma_t$'s, the quantity needed to describe their embedding in $\mathcal{M}$ is the second fundamental form, i.e. the extrinsic curvature $K_{ij}$, defined as 
\be
K_{ij} = \frac{1}{2}\pounds_{n} h_{ij} = \frac{1}{2N} \left(\partial_{t} h_{ij} - D_{i} N_{j} - D_{j} N_{i}\right)\,, \label{Lieder}
\ee
with $\pounds_n$ the Lie derivative along the direction $n^\alpha$.

That the quantities $(N,N^i,h_{ij})$ manage to store the same amount of information as $g_{\mu\nu}$ is a trivial result; as seen in Sect.~\ref{gravdofs}, the metric sports $10$ free functions (before any other argument is used to trim down the figure); in the ADM formalism, the lapse $N$ accounts for $1$ d.o.f., the shift $N^i$ adds three more, and $h_{ij}$ accounts for the remaining $6$ (it is represented by a $3\times3$ symmetric matrix). The actual value of $2$ ``true'' d.o.f.'s of GR resurfaces after introducing all the constraints given by symmetries (such as the spatial diffeomorphisms now emerging on each leaves), gauge fixings, etc.

\subsubsection{First-order reduction {\`a} la Ostrogradski}

To set the spotlight on the main aspects of the Hamiltonian formalism in ETG's, we focus now on general quadratic theories, as they are an excellent test-bench. The prototype for such models is still the action given in Eq.~\eqref{fRiem} of Sect.~\ref{linearization}, but for our purposes it is better to rewrite that action in a slightly different fashion; first, we re-introduce a cosmological constant-term, then make use of the Gauss--Bonnet invariant $\mathcal{G}$ to get rid of the Riemann-squared term, and further massage the result to arrive at the final action functional
\be
\mathcal{S}_{\text{quadr-var}} = \int_{\mathcal{M}} \sqrt{-g}\, \mathrm{d}^{4}x \left[\Lambda+\frac{1}{2\kappa}R-\frac{\alpha}{4} C_{\mu\nu\rho\sigma}C^{\mu\nu\rho\sigma}+\frac{\beta}{8} R^{2}\right] \,,\label{quadrvar}
\ee
with $\alpha,\beta$ two new coupling constants. In the expression above, we have neglected for the moment any boundary term required to counter-balance the uncompensated variations, as such terms demand a separate discussion --- see Sect.~\ref{bordohamil} below. The action~\eqref{quadrvar} must then be rewritten in ADM-form; by using the Gauss--Codazzi--Ricci relations~\cite{Wald:1984rg,Kluson:2013hza}, one can decompose the $4$-D Riemann tensor in its components tangent and normal to the leaves of the foliation. The final expression reads 
\be
\begin{split}
\mathcal{S}_{\text{quadr-ADM}} = \int \mathrm{d} t &\int_{\Sigma_{t}} \mathrm{d}^{3} y\, N \sqrt{h}\, \left\{\Lambda + \frac{1}{2 \kappa} \left[ \mathcop{R} + K_{ij} K^{ij} - K^{2} \right]\right.\\
&\left.- 2 \alpha \mathcop{C}_{i\bold{n} j\bold{n}} {{{\mathcop{C}^{i}}_{\bold{n}}}^{j}}_{\bold{n}} + \alpha \mathcop{C}_{ij\ell \bold{n}} \mathcop{C}^{ij\ell}_{\bold{n}} + \frac{\beta}{8} R^{2}\right\} \,,\label{quadradm}
\end{split}
\ee
where we have introduced the ``script'' letters variously denoting geometric objects built out of $h_{ij}$, the spatial covariant derivative $D_k$, and $\mathcal{T}_{\bold{n}}\equiv \mathcal{T}_{\mu}n^{\mu}$ --- we refer to~\cite{Kluson:2013hza} for the full details.

In the last reformulation given by Eq.~\eqref{quadradm}, a general quadratic ETG will contain second derivatives along the $t$-direction of $h_{ij}$ in the action, i.e. $\partial^{(2)}_t h_{ij} \equiv \ddot{h}_{ij}$. As we have just seen, to apply the Hamiltonian analysis we first have to recast the Lagrangian $\mathcal{L} (q,\dot{q},\ddot{q})$ into a form where at most first-order time derivatives are present, so that we are left with a $\mathcal{L} (Q^1,Q^2,\dot{Q}^2)$, which can be treated properly.\footnote{In general, other inequivalent choices of the evolution parameter are possible, even though this ``gauge'' choice does not help relieving the issue with the order of derivations; we shall come back on this matter in the following Sect.~\ref{bordohamil}.}

So to do, we have to re-absorb the undesired first-order time derivatives $\dot{h}_{ij}$ into a set of new variables, so that $\ddot{h}_{ij}$ becomes the first derivative of the mentioned new variable, and we are back on the track. This is a result one can achieve in many different ways; a solution proven effective is to pick the components of the extrinsic curvature tensor $K_{ij}$ as a new set of variables (this can always be done, as both tensors can be represented by $(3\times3)$ symmetric matrices), and later enforce the relation between $\dot{h}_{ij}$ and $K_{ij}$ --- namely, Eq.~\eqref{Lieder} --- as a constraint equation.

The new ensemble of canonical coordinates then becomes $(N,N^{i},h_{ij},K_{ij})$, with the subsequent determination of the conjugate momenta $(\pi^{0},\pi^{i},\pi^{ij},\varPi^{ij})$. We notice that, as it occurs in GR, $\dot{N}$ and $\dot{N}^{i}$ do not appear explicitly in the Lagrangian, and can be considered as cyclic variables; their conjugate momenta, $\pi^{0},\pi^{i}$, are then primary constraints.

Also, speaking of d.o.f.'s, we have that the newly-introduced ``Lagrangian coordinate'' $K_{ij}$ can sport at most $6$ new dynamical variables, making the possible number of d.o.f.'s for the quadratic theory~\eqref{quadrvar} top at $8$. Indeed, as we are about to discuss, the actual numbers are far smaller, as the growing of the canonical coordinates in Eq.~\eqref{dirac} is somewhat compensated by a corresponding growth in the number of first- and second-class constraints.

\subsubsection{Case studies, annotated}

So far, the description of the procedure has been completely general, and largely independent of the details of the action. At this point, however, we reach the fork in the path where the specific values of the parameters $\alpha,\beta,\kappa$ in Eq.~\eqref{quadrvar}, or~\eqref{quadradm}, begin to matter significantly in the scenario, so that different theories demand separate recipes. The calculations quickly escalate, to the point of becoming almost unmanageable, and many subtleties need be taken into account. While the common goal is to identify all the numbers required to calculate the result of Dirac's formula~\eqref{dirac}, each case has its own story.

In the following list we limit the discussion to a few relevant items, highlighting the agreement with what has been previously obtained by means of other extraction methods, and trying to deliver at least a vague flavour of the niceties of the protocol. One difference with the outcome of e.g. Sect.~\ref{linearization} is that here we allow also for the general presence of a cosmological constant, and we consider the possibility of the disappearance of all the scalar-curvature contributions. We refer to Refs.~\cite{Kluson:2013hza,buchbinder1992effective} for all the step-by-step calculations.
\begin{itemize}
\item $\alpha=\beta=0$. This is just the GR-case, so it is instructive to review it. The number $N$ in Eq.~\eqref{dirac} is given by the required initial Lagrangian coordinates, namely $10$. There are $8$ first-class constraints, i.e. $4$ given by the spatial and temporal diffeomorphisms (``momentum'' and ``Hamiltonian'' constraints), plus other $4$ from the cyclic character of lapse and shift. As a result, we are left with $2$ gravitational d.o.f.'s, as expected.
\item $\alpha=0,\; \beta\neq 0 $. The resulting ETG is nothing but $f(R)$-gravity. In this case, $N$ grows by $1$ with respect to GR,but no new constraints are added, nor additional symmetries. The same arguments as above apply, and we find one additional d.o.f., so that $\mathcal{N}=3$.
\item $\alpha\neq 0, \; \beta\neq 0$. In this case we are dealing with a total of $16$ Lagrangian coordinates; once again, the number of first-class constraints does not fluctuate with respect to GR, and no new second-class constraints arise. The counting of the d.o.f.'s stops at $8$, of which $6$ can be ascribed to fields other than the massless spin-$2$ graviton. In particular, around Minkowski background, the supplementary d.o.f.'s can be represented as a scalar field, plus a massive spin-$2$ Weyl poltergeist.
\item $\Lambda=0, \; \kappa^{-1}=\beta=0, \; \alpha\neq 0$. This is an interesting case, and one left unattended so far; the resulting ETG is commonly known as \emph{Weyl gravity}, the name emerging from the sole presence of Weyl-squared terms in the action. As a consequence, it is possible to prove that the action, and the resulting field equations, are conformally invariant, i.e. they are invariant under the reparametrization $g_{\mu\nu} \to \tilde{g}_{\mu\nu} = \phi^{2} g_{\mu\nu}$. Such theory is particularly appealing because the number of constraints is equal to the number of unstable directions in phase space --- we shall have more to say on this at the end of the Section. The Hamiltonian analysis starts once again with $N=16$, but now there are two more first-class constraints we must consider: one emerging from the mentioned conformal invariance, and the other due to the tracelessness of the momentum $\varPi^{ij}$ conjugate to $K_{ij}$. No second-class constraints emerge, and the overall sum gives a final figure of $6$ gravitational d.o.f.'s.
\item $\Lambda\neq 0, \; \kappa^{-1}=\beta=0, \; \alpha\neq 0$. This case amounts to Weyl gravity with a nonvanishing cosmological constant, which leads to a breaking of conformal invariance. Notwithstanding this new physical ingredient, early studies~\cite{buchbinder1992effective,Buchbinder:1987vp} seemed to concur on the persistence of $6$ gravitational d.o.f.'s at the end of the Hamiltonian analysis. More recent contributions~\cite{Kluson:2013hza}, however, show that one constraint had been erroneously considered and, hence, the actual number of physical d.o.f.'s is $5$.\footnote{At any rate, it has been ultimately established~\cite{Kluson:2013hza} that a Weyl-squared-plus-$\Lambda$ ETG is physically flawed, and its Hamiltonian structure suffers from critical --- and fatal --- illness.}
\item $\kappa^{-1}\neq 0,\;\beta=0, \; \alpha\neq 0$. In this last case we get an action made with an Einstein--Hilbert term plus Weyl gravity, as seen in Sect.~\ref{quadgrav}. The conformal invariance and the tracelessness mentioned above are broken at the level of the first-class constraints, but resurface as second-class ones, thus modifying the sum in Dirac's formula. As a net result, the number of d.o.f.'s becomes $7$, and on Minkowsky spacetime they can be encoded into one massless graviton, plus a massive spin-$2$ Weyl poltergeist. The further introduction of a $\Lambda$-term does not affect this argument, and does not change the final figure just obtained.
\end{itemize}  


As a concluding remark, we report another interesting feature of Weyl gravity which deserves a few comments~\cite{Maldacena:2011mk}. Around flat spacetime, the $6$ d.o.f.'s can be represented as one massless graviton, one massless spin-2 field --- the Weyl \emph{poltergeist} --- and a massless spin-1 field~\cite{Riegert:1984hf}. On backgrounds with constant, yet non-vanishing, curvature, apart from the massless graviton appears what is known as a ``partially massless spin-$2$ field''~\cite{Lee:1982cp}, i.e. a spin-$2$ object whose helicity can only assume the four values $(\pm 2,\pm 1)$, instead of the usual five $(\pm 2,\pm 1, 0)$.

\subsubsection{Auxiliary fields methodology}

Already at the level of quadratic corrections to GR --- in fact, a minimal extension --- the calculations required for the Hamiltonian analysis quickly become almost unmanageable, with a plethora of cross-checks needed to make sure one is not neglecting any piece, nor it is counting some twice. The problem arises, then, of finding whether it could be possible to extend the protocol to more complex higher-order ETG's.

A few results have been presented in Refs.~\cite{Deruelle:2009zk,Sendouda:2011hq}, starting however from a slightly different standpoint. We briefly review the main consequences here. Once again, one begins with a very general form of the action, namely 
\be
\mathcal{S} = \frac{1}{2 \kappa}\int_{\mathcal{M}}\sqrt{-g}\, \mathrm{d} x^4 f(g_{\mu\nu}, R_{\mu\nu\rho\sigma}) \,,\label{hamilaux}
\ee 
where $f$ is any sufficiently regular scalar function. The key preliminary step is the reduction of the higher-derivative action to one involving at most first-order derivatives of the dynamical variables, so that the final equations of motion involve second-order time derivatives at most. This is accomplished by introducing two new auxiliary fields, call them $\phi_{\mu\nu\rho\sigma}$ and $\rho_{\mu\nu\rho\sigma}$, as follows
\be
\mathcal{S}=\frac{1}{2 \kappa}\int_{\mathcal{M}}\sqrt{-g} \, \mathrm{d} x^4 \left[\phi^{\mu\nu\rho\sigma} \left(R_{\mu\nu\rho\sigma} - \rho_{\mu\nu\rho\sigma}\right) + f \left(\rho_{\mu\nu\rho\sigma},g_{\mu\nu}\right)\right] \,.
\ee 
The action above is dynamically equivalent to the one in Eq.~\eqref{hamilaux} as soon as one assumes that $\phi_{\mu\nu\rho\sigma}$ and $\rho_{\mu\nu\rho\sigma}$ are independent fields, so that the variation of the action yields the same set of field equations as before --- notice that, in the equation above, $\phi_{\mu\nu\rho\sigma}$ acts as a Lagrange multiplier, and we are postponing any discussion of the regularity and/or invertibility condition to the very end of the analysis, so to avoid all the branching issues germane to the introduction of auxiliary fields. 

The reformulation above differs from the protocol {\`a} la Ostrogradski, but ends up achieving the same core result, i.e. the reduction of the action to a strictly-first-order one, and hence from this point onwards the Hamiltonian analysis can follow the same conceptual lines as in the previous Section. By rewriting all the terms in an aptly-chosen ADM-form, it is possible to start counting all the constraints arising, which will of course depend heavily on the specific form of the function $f$. This variation on the Hamiltonian theme might appear less straightforward from the point of view of the choice of the canonical variables, but in the end it works as fine. 

An interesting aspect of adopting this strategy --- which, by the way, is not limited to quadratic cases, but can in principle work with arbitrary ETG's --- is that the ADM-form of the action allows us to see immediately that any gravitational d.o.f.'s besides those collected under the graviton will automatically be encoded in the tensor
\be
\psi^{ij} = - 2 h^{ik} h^{jl} n^\mu n^\nu \phi_{k \mu l \nu},
\ee  
The tensor $\psi^{ij}$ is purely spatial and symmetric, and hence can carry at most 6 d.o.f'.s. Which is a neat and simple explanation of the splitting of the extra d.o.f.'s in a spin-$2$ and a spin-$0$ part: this is in fact the most straightforward and convenient way in which $\psi^{ij}$ can be decomposed; the trace $\psi^i_i$, a single scalar, accounts for the spin-$0$ component, whereas the traceless (``deviatoric') part stores the other $5$ d.o.f.'s in the compact form of a spin-$2$ poltergeist.

We stress, however, that the statements above only constitute an upper bound for the actual number of extra d.o.f.'s; the true figure emerging from Dirac's formula~\eqref{dirac} might be smaller, depending on the specific nature of the theory at hand (and its symmetries and gauge redundancies).

\subsection{Considerations on the method}

Upon looking at the remarkable results achieved by using the Hamiltonian analysis for ETG's, a few conclusions emerge naturally. First and foremost, this method is completely background-independent, and as such it makes for a serious improvement over all the other protocols reviewed previously. Sure, the calculations can become almost unmanageable, but they guarantee a full understanding of the dynamics and physical content of a given model.

Also, as it emerges from Dirac's formula, the number of ``true'' gravitational d.o.f.'s is ultimately confirmed to be a fundamental, non-negotiable property of every theory under scrutiny, hence the possible ``disappearance'' or ``cloaking'' of dynamical variables in specific spacetime configurations must be strictly regarded as accidental, associated to the additional symmetries of the solutions, rather than to the structure of the field equations.

At the same time, the method confirms as well that the \emph{representation} of the d.o.f.'s, i.e. their arrangement as components of geometric quantities defined over the spacetime manifold, tends to heavily fluctuate, and can be accommodated in many inequivalent ways, ranging from sets of scalars to complex, partially-massless poltergeists. Indeed, as already stressed in the Introduction, the particle content, or rather the classification of the propagating fields in terms of spin and mass, is a background-dependent notion, and one far less fundamental than the dynamical content itself, notwithstanding the high regard such notion enjoys.

Another aspect to be further highlighted is that, in the Hamiltonian context, many of the advantages of the method can become sources of misinterpretations and subtle mistakes: the gauge fixings, the choice of the new Lagrangian coordinates in the order-reduction {\`a} la Ostrogradski, the classification of the constraints: it all adds up to the generally high level of difficulty of the protocol, both computational and interpretative.\footnote{These subtleties can result in macroscopic modifications: in a few moments we shall address the issue of the boundary terms, and see how trading one set of canonical coordinates for another severely affects the surface counter-terms needed to render the variational problem for the ETG's at hand well-posed.}

Finally, it is worth noticing that the Hamiltonian standpoint is not the only chance one has to address the problem of the d.o.f.'s in a completely general setting: the search for the constraints entering Dirac's formula can be also conducted in a fully-covariant fashion, as done in~\cite{LeeWaldcf}, or by adopting a modification of the Lagrangian formalism lying somewhere between the canonical and covariant ones~\cite{lagform}.

\subsection{Supplementary material}

For the concluding paragraphs of this section, we present a few remarks connected to the general topic of the Hamiltonian formalism, each one possibly of interest per se. We begin with a short discussion of the behaviour of the boundary terms in ETG's, when the former are rewritten in the language of the canonical variables; then, we elaborate more on the issue with Ostrogradski's instability, and see whether there are ways out (and how to implement them). Finally, we recollect an old argument, originally due to Einstein himself, which might shine a whole new light on the structure of field theories, and in particular on the family tree of gravitational models.

\subsubsection{Boundary terms: the Hamiltonian standpoint(s) \label{bordohamil}}

While the Hamiltonian formalism qualifies as the most straightforward tool in the search for the degrees of freedom of a physical theory, it severely affects the construction of the boundary terms making the variational problem well-posed. One gets to the point that surface pieces can be made appear and disappear, depending on the specific choice of the dynamical variables.

Such conclusion has been highlighted in a recent paper~\cite{Kluson:2013hza} on quadratic gravity, devoted to a wide analysis of the properties of ETG's characterised by the action
\begin{equation}
\mathcal{S} = \int_{\mathcal{M}} \sqrt{-g} \, \mathrm{d}^4 x \left[ \Lambda + \frac{R}{2\kappa} - \frac{\alpha}{4} C_{\mu\nu\rho\sigma} C^{\mu\nu\rho\sigma} + \frac{\beta}{8} R^2 + \gamma \mathcal{G} \right] \;, \label{eq:klusonaction}
\end{equation}

The definition of the Weyl tensor and its symmetry properties allow to find the following equivalence
\begin{equation}
C_{\xi\zeta\rho\sigma} C^{\xi\zeta\rho\sigma} = 2 \left( R_{\mu\nu} R^{\mu\nu} - \frac{R^2}{3} \right) + \mathcal{G} \;,
\end{equation}
so that the action in Eq.~\eqref{eq:klusonaction} can be rewritten equivalently as
\begin{equation}
\mathcal{S} = \int_{\mathcal{M}} \sqrt{-g} \, \mathrm{d}^4 x \left[ \Lambda + \frac{R}{2} - \frac{\alpha}{2} R_{\mu\nu} R^{\mu\nu} + \left( \frac{\alpha}{6} + \frac{\beta}{8} \right) R^2 + \left( \gamma - \frac{\alpha}{4} \right) \mathcal{G} \right] \;. \label{eq:klusonaction2}
\end{equation}
The two expressions~\eqref{eq:klusonaction} and~\eqref{eq:klusonaction2} differ by a multiple of the Gauss--Bonnet term, which by topological considerations has no consequence on the resulting field equations. The two theories are therefore dynamically equivalent, and sport the very same number and type of gravitational d.o.f.'s. When one enforces the Hamiltonian analysis of the ETG, it is a matter of choice which one of the two actions to consider, together with the general issue of picking a set of canonical variables. Which leads to two sorts of issues.

\paragraph{Choice of the dynamical variables} During the implementation of the required ADM, foliation-adapted coordinates, there is a residual ambiguity regarding the notion of time differentiation. Indeed, one might make use of various timelike vectors --- e.g. $(\partial/\partial t)^\mu$, or $m^\mu = N n^\mu$, and so forth --- to identify the direction along which the ``time'' flows in the ADM decomposition. Such problem is mirrored by the independence of the Lagrangian from $N^i$, which is a cyclic coordinate resulting in a first-class constraint. The physics remains unaffected by the different choices; not quite so the appearance of all the expressions and calculations.

With all these caveats in mind, one can nonetheless formulate the Hamiltonian counterpart of an ETG, rewriting the action (both the bulk and surface part) in terms of the chosen ADM variables and additional gravitational d.o.f.'s. Notice, however, that in the new, first-order representation of the given theory, the vanishing variations on the boundary $\partial \mathcal{M}$ will be those corresponding to \emph{all} the dynamical variables, hence typically it will be $\delta h_{ab} = \delta K_{ab} = 0$ on $\partial \mathcal{M}$.

\paragraph{Issues with the boundary terms} In the case of quadratic gravity, a long yet almost straightforward calculation shows that all the boundary terms on the foliated spacetime reduce to pieces which are linear in the ADM variables and in $K_{ab}$~\cite{Kluson:2013hza}. This has the powerful consequence that, independently of the value of the coefficients $\alpha,  \beta, \gamma$ in the action~\eqref{eq:klusonaction2}, all boundary terms will vanish identically once varied --- their arguments are linear functions of $\delta h_{ab}$ and $\delta K_{ab}$, which are zero on $\partial \mathcal{M}$. The only surviving boundary terms are the ADM-translated Gibbons--Hawking--York blocks, which will reabsorb the uncompensated surface integrals emerging from the variation of the bulk action, as expected.

While this result is germane to the whole class of quadratic gravity theories, it has to be stressed that it holds only for one specific choice of dynamical variables, i.e. when one collects the gravitational d.o.f.'s within the pair ``ADM variables \emph{\&{}} extrinsic curvature''. Even if this is a typical choice, nothing prevents one from performing a canonical (non-linear) transformation and switch to a different set of variables, call them $(H_{\mu\nu},k_{\mu\nu})$. As soon as one does so, however, the boundary terms cannot be written anymore as linear combinations of the $H_{\mu\nu}$'s and $k_{\mu\nu}$'s, and hence do not vanish anymore under variation; sure, the variations $\delta H_{\mu\nu}$ and $\delta k_{\mu\nu}$ will again be identically zero on $\partial \mathcal{M}$, but new surface terms will crop up, and require apt handling care to be dealt with.

\subsubsection{A note on Ostrogradsky's (in)stability\label{ghosts}}

We have stressed that higher-derivative theories are in general unstable, as can be proved by application of a noteworthy theorem by Ostrogradski~\cite{Woodard:2006nt,Eliezer:1989cr}. The reason is that in such models the canonical Hamiltonian turns out to be linear in the momenta, and then unbounded from both above and below\footnote{See however \cite{Kaparulin2014} for a different perspective on the problem at the classical and quantum level.}.

At the classical level, instabilities in a field theory are often associated to \emph{ghost} fields, i.e. fields which have negative kinetic energy~\cite{Sbisa:2014pzo}. In a quantum context (e.g., when the gravitational field excitations on a maximally symmetric spacetime are quantised), such ghosts can be seen either as states with negative energy --- which results in an instantaneous decay of the vacuum as soon as Lorentz invariance is implemented~\cite{Eliezer:1989cr} --- or as states with negative norm, which implies that the underlying theory is non-unitary, and therefore non-predictable in its dynamical evolution~\cite{Hindawi:1995an,Stelle:1976gc,Stelle:1977ry,Chiba:2005nz,Nunez:2004ts}.

Ostrogrdski's theorem, however, is based on the crucial assumption that the starting Lagrangian function is regular, which is precisely \emph{not} the case of the ETG's, which in turn are characterised by singular Lagrangians. Consequently, a straightforward evocation of Ostrogradski's theorem makes little sense in this case. As a result, the issue with the instability of ETG's can be relieved, at least in some specific cases. For instance, fourth-order $f(R)$-theories can be made stable in spite of their higher-derivative character: the catch is that $f (R)$-theories present at once one, and one only, unstable direction in their phase space, but also one local constraint, which aptly tames and cures the instability. For a detailed and clear analysis of this case, including a thorough discussion of subtleties and misconceptions, see~\cite{Woodard:2006nt}.

More generally, the first-class/gauge constraints can restrict the dynamics of a given theory in such a way that the regions of the phase space where instabilities arise become forbidden; the resulting model is thus free of problems. For this to happen, the number of gauge constraints must be at least equal to the number of unstable directions in the phase space~\cite{Kluson:2013hza,Woodard:2006nt}). Weyl gravity, with Lagrangian $\sqrt{-g}\, C^{\alpha\beta\gamma\delta} C_{\alpha\beta\gamma\delta}$ is another example in this sense: the number of constraints matches the number of unstable directions in phase space, therefore such theory can avoid Ostrogradski's instabilities, at least in principle~\cite{Kluson:2013hza}. Upon noticing this fact, the community has started introducing extra constraints for originally unstable ETG's, hoping to tame them one by one (see e.g.~\cite{Chen:2012au} and references therein).

Another way out of Ostrogradski's instabilities lies within the Effective Field Theory (EFT) paradigm. At the core of such research program lies the idea that the higher-derivative Lagrangians emerging in ETG's are just approximate truncations of a complete theory, for which the full, yet unknown, form could reabsorb the instabilities and tame them in an appropriate way~\cite{Yunes:2013dva,Burgess:2003jk}.

Finally, in a similar, and possibly connected fashion, also non-local theories might provide better answers to the issue of instabilities. It should be noted that Ostrogradski's construction can indeed fail for theories containing infinitely many derivatives \cite{Barnaby:2007ve}. In fact, non-local ETG's and field theories have attracted more and more attention lately \cite{Dirian:2014bma,
Biswas:2013kla,Maggiore:2014sia,Dirian:2016puz,Calcagni:2014vxa,Li:2015bqa,Giaccari:2016kzy}, and the study of their constraints in the Hamiltonian formalism is considered promising as per the handling of classical and semi-classical instabilities~\cite{Woodard:2006nt,Eliezer:1989cr,Simon:1990ic}. Also, truncated non-local models often reduce to higher-derivative theories of the EFT type, with an interesting link between the two paradigms.

\subsubsection{Einsteinian ``strength'' and ETG's \label{forza}}

As a fair conclusion to this Section, we deem it interesting to present yet another way to calculate at least the number of degrees of freedom in any field theory for which the dynamical equations are known. The method dates back to the Fifties, and was originally conceived by Einstein himself~ \cite{einstein2001meaning}, in an effort to prove once again the effectiveness and simplicity of his relativistic theory over other models.

The idea somehow sidesteps the usual considerations in field theories, and stems from the concept of ``strength'' of a system of field equations, i.e. the number of free parameters in a given model, once all the field equations and their symmetries have been introduced to determine the values of the parameters. In this context, a ``strong'' theory is one where all the values of the parameters are locked by the dynamical equations, whereas a ``weaker'' theory allows for a more or less large degree of arbitrariness.\footnote{For a broader perspective on this issue, and its potential connection to other relevant epistemological questions in modern Physics, see the recent preprint~\cite{Barrow:2015fga}.}

It has been later found~\cite{:/content/aip/journal/jmp/15/4/10.1063/1.1666669,:/content/aip/journal/jmp/16/4/10.1063/1.522619,Kaparulin2013} that Einstein's idea can be connected to a method looking for the number of degrees of freedom in a given ETG and field theory in general, with other interesting links to the issue of determining the required Cauchy data~ \cite{1988AN....309..357M}. A more gravity-oriented research, with specific applications to ETG's, has been pursued in~\cite{Garecki:2002xc,Garecki:2003ri}.

While the concept of the ``strength'' of a field theory does not parallel the power of the Hamiltonian formalism, it remains a powerful (and way much faster) acid test to find out the number of dynamical variables. Also, and this was of no secondary importance at the time Einstein came up with the idea, this protocol works flawlessly even when a field theory cannot be expressed in a Lagrangian form, but only via its field equations~\cite{1988AN....309..357M}.

We now briefly summarise the method, following~\cite{:/content/aip/journal/jmp/16/4/10.1063/1.522619,Garecki:2002xc}. We start with a general analytic field function $\Phi$, depending on $d$ variables $y^\alpha$, $\alpha = 1, \dots ,d$ (since we are concerned with ETG's we can think of $d$ as the dimensionality of spacetime, and the variables $y^\alpha$ as coordinates on a Riemannian manifold with a Lorentzian metric). We can Taylor-expand $\Phi$ in a power series around a ``point'' $\bar{y}^\alpha$; each term in the expansion will sport different coefficients $c_k$, given by the $k$-th derivatives of the function evaluated at $\bar{y}^\alpha$. It is possible to prove that every function $\Phi$ will be uniquely characterised by the set of the $c_k$ coefficients in its Taylor series, and vice-versa.

Now, it is manifest that, if $\Phi$ is completely unconstrained, then not a single $c_k$ in the expansion is known, or can be calculated in some way; the opposite is also true: if all the $c_k$'s are arbitrary, then the behaviour of $\Phi$ cannot be read out in any way, and the function is unconstrained. If, instead, the function must satisfy some set of field equations and/or constraints, then the number of free coefficients in the Taylor expansion will decrease, leaving only a certain number of them arbitrary. The concept of ``Einstenian strength'' of a system of equations builds upon this result.

What Einstein proposes is to Taylor-expand a function $\Phi$ obeying a set of field equations, and count the number of coefficients of order $n$ left free after all the possible relations emerging from the equations of motion and from the gauge freedom are used to frame the values of the said coefficients. The number $Z_n$ of free terms of order $n$ can be always written as 
\begin{equation}\label{schutz}
Z_{n}=\sum_{k=1}^{d} N_{k} \stre{k}{n} \;,
\end{equation}
where the symbol $\tstre{k}{n}$ is a shorthand notation for the combinatorial structure
\begin{equation}
\stre{k}{n} \equiv  \frac{(n+k-1)!}{n!(k-1)!}
\end{equation}
The number $N_{k}$ in Eq.~\eqref{schutz} represents the number of free functions of $k$ variables in the solution. Notice that, for a physically sound theory, we expect that for $k=d$ it is $N_d = 0$ identically, as $N_d$ represents the number of functions of $d$ variables left free (i.e., completely unrestricted) by the field equations. 

An estimate of the ``strength'' of a field theory can be retrieved by calculating the large-$n$ limit of the ratio $Z_{n}/\tstre{k}{n}$, and extracting the coefficient of the $1/n$ term. In~\cite{:/content/aip/journal/jmp/16/4/10.1063/1.522619} it was first realized that the Einsteinian strength of a system of field equations is closely related to the number of dof's. 

For our case of ETG's, we are interested in determining the number of degrees of freedom of the model; to do so, we look for the number of Cauchy data on a $(d-1)$-dimensional hypersurface on the spacetime manifold, which is exactly twice the number of true degrees of freedom for the theory. This implies that we have to evaluate the number $N_{d-1}$ in \eqref{schutz}. A small value for $N_{d-1}$ will imply a ``stronger'' theory, whereas a large value of the coefficient will signal a ``weaker'' character of the resulting ETG.

A few examples will make the meaning of the previous statement clearer. Let us consider first a generic, purely metric, quadratic theory for which $d = 4$, as in Eq.~\eqref{aquad}. Without considering the field equations, the number of free parameters in the symmetric tensor $g_{ab}$ is $10$; at order $n$, the total number of coefficients is thus $10 \cdot \tstre{4}{n}$.

We can now play with the symmetries and the field equations. General covariance, the main gauge freedom, implies $4 \cdot \tstre{4}{n-1}$ relations between $n$-th order coefficients. We have $10$ equations of motion of $4$-th order (in general), thus we obtain other $10 \cdot \tstre{4}{n-4}$ relations between the $n$-th order coefficients --- we have to derive the field equations $n-4$ times to get such relations for the $n$-th order coefficients. Finally, the field equations satisfy the $4$ differential Bianchi identities, which themselves are of $5$-th order, and hence give $4 \cdot \tstre{4}{n-5}$ relations among the equations. This means that not all the relations generated by the field equations are independent.

The number of free coefficients is then given by the sum
\begin{equation}
Z_{n} = 10 \stre{4}{n} - 4 \stre{4}{n+1} - \left\{10 \stre{4}{n-4} - 4 \stre{4}{n-5} \right\} \;,
\end{equation} 
and we are interested in the number $N_{3}$, extracted after the previous line is arranged so as to match Eq.~\eqref{schutz}. The result is $N_{3} = 16$, which gives twice the number of actual degrees of freedom for a general quadratic ETG, as expected.

As a second case-study, we focus on pure Weyl gravity, with Lagrangian density equal to $C^{\alpha\beta\gamma\delta} C_{\alpha\beta\gamma\delta}$. By following the protocol, one obtains the expression~\cite{1988AN....309..357M}
\begin{equation}
Z_{n} = 10 \stre{4}{n} - 4 \stre{4}{n-1} -\left\{10 \stre{4}{n-4} - 4 \stre{4}{n-5} + 1 \stre{4}{n} - 1 \stre{4}{n-4} \right\},
\end{equation} 
where the last two terms come from two additional ingredient of this ETG: its conformal invariance, and the vanishing of the trace in the field equations. As can be seen, in this case $N_{3}=12$. Thus, conformal invariance helps removing some coefficients, whereas instead the tracelessness of the equations of motion renders some of the relations no more independent, hence increasing the number of free coefficients. 

By this brief recollection of the application of Einstein's concept of ``strength'' it is easy to spot all its power and limits. The main drawback is that it gives nothing more than a mere number, with no further information on how the dynamical variables can be later reshuffled to form more complex geometrical objects (vectors, tensors, spinors, bi-tensors, etc.). Also, it relies heavily on the knowledge of the field equations and, a priori, of all the possible symmetries of the theory, which might be a tough riddle to solve without having the Lagrangian function at hand.

Yet, as remarked and presented, the protocol allows for a drastically faster way to count the actual degrees of freedom in a field theory via its Cauchy data, and can be applied whenever one has an idea of a possible set of equations of motion --- see in this sense how the concept has been successfully used in the study of a traceless Bach--Weyl--Einstein theory, which does not even admit a variational formulation~\cite{1988AN....309..357M}.

The remarkable effectiveness of this method has generated a recent peak of interest, especially in the study of vast branches of the gravitational sector where additional geometric structures are involved, and the focus is mainly on the field equations --- scalar-tensor theories, Weintzb{\"o}ck's teleparallel gravity, ETG's with torsion or non-metricity, etc. In~\cite{Garecki:2002xc}, it is possible to find a collection of relevant results for quadratic ETG's of the purely metric, metric-compatible, and metric-affine type, where a variational formulation is sometimes missing, or unclear; there, the figure $74$ appears as an average for the number of d.o.f.'s typically involved in the reported scenarios. Such number substantially agrees, at least in terms of the orders of magnitude, with other estimates performed in the context of gauge theories of gravity~\cite{Hehl:2012pi}.

What might be interesting to see is whether this method could be applied to non-local ETG's as well: indeed, Einstein's idea works directly at the full non-linear level, and hence avoids any linearisation procedure around specific (and often, over-simplified) backgrounds. The concept of ``strength'' might then shine a light on two relevant aspects: the number of actual degrees of freedom of a non-local model, and its relation to the higher-derivative, truncated theories emerging in the EFT paradigm. While a full result could only come from other approaches, such as the Hamiltonian formalism, a rough estimate could already offer precious insights on where to look for.%
\section{Discussion and Conclusions \label{conclusions}}

The search for the relations linking different gravitational theories and the assessment of their physical equivalence remains a largely unexplored field, pursued sparsely by a community of committed enthusiasts. At the same time, we ought to admit that the recent wealth of discoveries in large-scale astrophysics and cosmology, and the theoretical quest for the ultimate quantum theory of gravity make for a powerful motivation for settling a deeper understanding of the family tree of gravitational phenomena, its innermost structure, and its most effective formulation.

In this work, we have focused on the issue of uncovering the actual degrees of freedom characterising a theory of gravity, and their relation to the field content --- or representations --- of the theory itself. A most crucial aspect, one on which we have periodically insisted throughout these pages, is that while the actual degrees of freedom of a theory are an \emph{intrinsic property} of the latter, the field content that embodies them, the actual field representation of the model, is mainly a \emph{conventional choice}, and it is not necessarily unique. In fact, we have seen that several higher-curvature ETG's can be recast as GR plus a variable amount of extra (non-minimally coupled) fields.

It is then tempting to conjecture that such reformulation in terms of pure-GR plus extra fields can be achieved for \emph{any} ETG. Such a result would not only lead to the immediate emergence of equivalence classes of models, but it would also allow us to assess very easily their physical viability, as most of the tools currently at hand to test alternatives to Einstein's model are designed to spot precisely the presence of ``non-metric'' contributions~\cite{will-rel}.

In the case of quadratic ETG's, this reformulation is indeed possible, and leads to the the well-known result that these theories are generically not viable because of the presence of unstable extra d.o.f.'s, taking the form of \emph{ghost} fields. In view of this conclusion, the analysis of the number and nature of the extra d.o.f.'s is obviously a crucial, required step one has to take long before trying to use such theories to explain observed phenomena as dark energy or dark matter.

At the same time, from what we have reviewed here it should be now clear that extracting and isolating the actual d.o.f.'s is far from a trivial task: most of the simpler methods so to do provide inconclusive results, the few noticeable exceptions teach very little about the underlying structural difficulties, and the state-of-the-art techniques are plagued by subtle pitfalls and computational fatigue making any tiny advance an almost-Sisyphean endeavour.

For sake of convenience, we briefly sum up the main findings reported in the previous Sections.

\paragraph{Boundary terms as telltale signs of extra d.o.f.'s} The metric variation of gravitational Lagrangians almost invariably generates uncompensated boundary terms. In a few cases, as in GR, these can be reabsorbed by aptly deploying some Gibbons--Hawking--York-like counter-terms, but in general their presence can be considered a definite signature of some deviation from the figure of $2$ associated to the number of d.o.f.'s in a purely metric scenario. At the same time, the examination of the surface structure can be at best a diagnostic tool, as it is in general unable to identify precisely the nature and form of these extra dynamical variables. The few exceptions where the method seems to work properly can be treated far more precisely using other techniques.

\paragraph{Particle content from field linearisation} A much more effective approach to uncover the extra d.o.f.'s starts by extracting the particle content of a theory via a linear expansion of the metric around some suitable background (Minkowski flat spacetime, or some
other maximally symmetric solution of the field equations). However, this (not uncommon) approach is far from being safe. Indeed, a short-sighted application of the recipe can expose only a fraction of the extra d.o.f.'s, because in some cases the high symmetry of the background actually ``freezes'' some fields, preventing their emergence at the sole linear level.

This result is often interpreted as the absence of any phenomenology related to these ``frozen'' d.o.f.'s in such spacetimes. Which is no more
than wishful thinking: any little step away from the highly symmetric backgrounds would easily ``reactivate'' the fields carrying the extra dynamical variables, with highly non-negligible consequences in view of the non-linear character of gravity. As a conclusion, the linearisation protocol can be deemed able to expose the actual d.o.f.'s of a given theory only depending on the representation that the latter induces on the background. Unfortunately, the representations associated to the most common MS spacetimes typically do not carry enough dynamical content to encode all the d.o.f.'s.

\paragraph{Auxiliary fields and change of representation} The auxiliary-fields method moves from the need to reduce the differential order of the field equations of a given theory to the familiar number of two; to do so, one has to introduce new fields and reshuffle the dynamical variables. This feature suggests an immediate application of the protocol to the class of higher-order ETG's. In the case of models quadratic in the curvature, a total of eight dynamical d.o.f.'s emerge, which are then encoded into a suitable number of spin-$0$ and spin-$2$ particles (no spin-$1$ fields are present). The spin-$2$ objects can be massless (the graviton, carrying $2$ d.o.f.'s), or massive (the Weyl \emph{poltergeist}, sporting $5$ d.o.f.'s, and always a ghost state). As a result, such theories can in principle be cast as standard GR plus a massive graviton and a scalar field (the latter accounting for the remaining single d.o.f.). Also, by fine-tuning the coefficients in the action, it is possible to conceive non-trivial Lagrangians (beyond the well-known case of $f(R)$ models) which are ghost-free and can be represented as general scalar-tensor theories of the Horndesky type.

Unfortunately, as soon as one goes beyond fourth-order theories and considers general $f(g_{\mu\nu},R_{\mu\nu\rho\sigma})$ Lagrangians, the auxiliary-fields method becomes rapidly unmanageable (for instance, there can be too many choices for the auxiliary variables, many of which of little to no physical significance) and only in very few cases a full analysis can be carried on.

\paragraph{Action expansion and effective quadratic theory} Another technique typically adopted suggests to expand the action of an ETG up to second order in curvature invariants, around maximally symmetric spacetime solutions. As the outcome is invariably a quadratic theory, the study of such ETG provides results which hold for a vast class of models up to second-order corrections. While this protocol is different from linearizing the \emph{metric} around a fixed background, it ends up exhibiting the same old problem: the expansion around some MS spacetime, in view of the high degree of symmetry in the background, is unable to fully expose the actual d.o.f.'s of a theory, for the associated induced representation (field content) inevitably ``freezes'' some of the fields. The non-dynamical character of some d.o.f.'s emerging from this procedure is, however, just a byproduct of an initially too-restrictive choice of the background, and is quickly reverted back as soon as more general spacetime solutions are introduced.

As a further remark, it is worth noticing that, even though our analysis was limited to a few particular classes of ETG's and mainly focused on MS solutions, our conclusions are far more general. For example, in the case of Ho\v{r}ava--Lifshitz theory,\footnote{Ho\v{r}ava--Lifshitz gravity is a higher-derivative, Lorentz-violating theory that is not fully diffeomorphism-invariant~\cite{Horava:2009uw,PhysRevD.79.084008}. We mention this model because our conclusions hold for Ho\v{r}ava--Lifshitz gravity as well, far outside the context of higher-order ETG's.} it was believed from studies at the level of linear perturbations around cosmological backgrounds~\cite{Gao:2009ht} that there were no extra d.o.f.'s with respect to those of standard GR. Later on~\cite{Blas:2009yd}, it has been realized that the theory possesses indeed an additional scalar d.o.f., which becomes singular for cosmological backgrounds (whence the initial difficulty to confirm its presence). An analogous situation arises in $f(R,\mathcal{G})$ theories \cite{DeFelice:2010hg}, in which a scalar d.o.f. is absent in the expansion around FRW spacetimes, but manifests itself in the anisotropic case. 

\paragraph{The Hamiltonian analysis, outpost of progress (with a caveat)} The ultimate extraction method to frame the number and dynamics of the gravitational d.o.f.'s could likely be provided by a full Hamiltonian analysis. The extreme effectiveness in assessing a dynamical content is a known feature of this strategy, and the method fares much better than any other available recipe. Unfortunately, the canonical approach, while impeccable at determining the number of d.o.f.'s --- thanks to the Dirac formula~\eqref{dirac} --- is technically challenging and highly non-trivial, and requires mastering both the canonical formalism and all the tricks to deal with boundary terms, order-reduction techniques, field redefinitions, and so forth. So far, only quadratic ETG's have been fully explored in this context, and not much more than a confirmation of what has been already found elsewhere has been extracted. There are some daring proposals tackling more complex scenarios (possibly with a little help from other techniques), but a completely general statement about the dynamical content and representation of ETG's is currently out of sight, even in the context of Hamiltonian analysis.

Also, just before the closing remarks, it is perhaps worth inserting a cautionary note about the physical significance of the results reported above, especially those regarding the Hamiltonian analysis. Indeed, despite its effectiveness and reliability when it comes to uncovering the extra physical d.o.f.'s of a theory, the last method is also somewhat blind to their physical relevance in the actual phenomenology we, as observers, might experience. This issue partly stems from the fact that all higher-order theories of gravity can be ultimately considered as effective field theories and, as such, they are limited in their range of validity, requiring some UV completion to overcome their intrinsic boundaries. In this sense, then, the mere existence of problematic extra degrees of freedom might even be safely considered \emph{harmless}, as long as their relevance and/or emergence does not occur on the specific background/physical context within which we are applying the theory (e.g. a cosmological setting, or a Post-Newtonian scenario).

Consequently, in spite of the power of the Hamiltonian analysis and all the techniques alike, one might still say that the most important physical check required to be performed consists in the phenomenological analysis of the theory in the sole context of specific, physically interesting background solutions --- which amounts to adopting the linearization recipes outlined above.

\medskip

In conclusion, the picture emerging from this investigation should convey the message that, to date, the analysis and characterisation of the actual degrees of freedom of a theory of gravity has to resort to a combination of procedures, as this seems to be the only way to to overcome the weaknesses of each single method, and at once enhance the merits of the available techniques. Also, it seems fair to say that all the current protocols are somewhat limited in their applicability, and work properly only on a relatively small sub-class of gravitational Lagrangians --- already in the comparatively narrow set of ``purely metric'' theories of gravity; the extension to e.g. metric-affine, affine and more general paradigms would open another Pandora's box.

At the same time, we would like to conclude on a brighter note, and stress that, even though this remains a critical review, we were more than happy to discover the vast amount of scattered results available in the literature, which altogether make quite an impressive milestone towards the finding of a satisfying ultimate answer. If some of the tools developed so far might now look somehow blunt, it is just because the challenge the community faces is harder than ever, and demands even smarter ideas, and more committed practitioners. We therefore hope that this comprehensive presentation could be used to foster and urge the search for the leap forward we deeply need.

The quest for the true nature of gravity will certainly pass through the understanding of the principles, structure, and interpretations of our models of it, and any improvement in the taxonomic studies might help unveiling the key to unlock the much sought-after treasure chest.

\begin{acknowledgments}
SL, AB and ML wish to acknowledge the John Templeton Foundation for the supporting grant \#51876. The opinions expressed in this publication are those of the authors and do not necessarily reflect the views of the John Templeton Foundation. AB acknowledges the support of the Austrian Academy of Sciences through Innovationsfonds \textit{Forschung, Wissenschaft und Gesellschaft} and the University of Vienna through the research platform TURIS. EDC gratefully acknowledges Prof. L.~Negrisin and the whole academic and administrative staff at Liceo ``Galilei'', Trieste, for the enlightening discussions, crucial support, and warm hospitality.
\end{acknowledgments}%
%

\end{document}